\documentclass[12pt]{article}

\usepackage[margin=1in]{geometry}
\usepackage{amsmath,amssymb}
\usepackage{newtxtext,newtxmath}                 
\usepackage{booktabs,threeparttable,longtable,array,multirow,makecell}
\usepackage{graphicx,float,caption,subcaption}
\usepackage{pdflscape}
\usepackage[natbibapa]{apacite}                  
\usepackage{setspace}
\usepackage[hidelinks]{hyperref}
\usepackage{xcolor}
\newcommand{\RunsPerStrike}{0.135}

\newcommand{\AgreeTwentyTwo}{91.4}
\newcommand{\AgreeTwentyThree}{91.6}
\newcommand{\AgreeTwentyFour}{98.6}

\newcommand{\AgreeTwentySix}{98.6}
\newcommand{\NPitches}{1,203,755}
\newcommand{\NCalled}{657,201}
\newcommand{\NPA}{308,959}
\newcommand{\CalibAgree}{98.6}
\newcommand{\AgreeABSHeight}{99.1}
\newcommand{\UpperThirdChange}{13.9}
\newcommand{\LowEdgeChange}{-15.2}
\newcommand{\EqInsideHumanF}{25.8}
\newcommand{\EqInsideABSF}{1.4}
\newcommand{\EqOutsideHumanF}{11.7}
\newcommand{\EqOutsideABSF}{3.2}
\newcommand{\EqInsideABSp}{0.23}
\newcommand{\EqOutsideABSp}{0.02}
\newcommand{\ExpoRHPRHB}{0.007}

\newcommand{\ExpoLHPLHB}{0.043}
\newcommand{\ZoneTopTallHuman}{105}
\newcommand{\ZoneTopTallABS}{110}
\newcommand{\ZoneTopShortHuman}{102}

\newcommand{\StarPost}{1.1}
\newcommand{\StarYrTwentyTwo}{-1.15}
\newcommand{\StarYrTwentyThree}{-1.75}
\newcommand{\StarYrTwentyFour}{-0.43}
\newcommand{\StarYrTwentyFive}{-0.35}
\newcommand{\StarYrTwentySix}{-0.17}
\newcommand{\CountAheadPre}{6.1}
\newcommand{\CountTwoPre}{-6.7}
\newcommand{\CountTwoPreAbs}{6.7}
\newcommand{\FrameSDPre}{0.93}
\newcommand{\FrameSDPost}{0.14}
\newcommand{\FrameRelMin}{0.60}
\newcommand{\FrameRelMax}{0.81}
\newcommand{\FrameRunsPre}{4.5}
\newcommand{\FrameRunsPost}{0.7}
\newcommand{\BetaAllTwentyFour}{0.047}
\newcommand{\BetaAllTwentyThree}{0.030}
\newcommand{\BetaAllTwentyTwo}{0.001}
\newcommand{\AccelAll}{0.018}
\newcommand{\AccelLHP}{0.058}
\newcommand{\AccelRHP}{0.005}
\newcommand{\AccelLHPbal}{0.052}
\newcommand{\BetaLHPTwentyFour}{0.060}
\newcommand{\BetaRHPTwentyThree}{0.034}
\newcommand{\FSPooled}{0.75}
\newcommand{\FSF}{49}
\newcommand{\RFTwoStrikeTenCm}{0.57}
\newcommand{\RFThreeBallTenCm}{-0.34}
\newcommand{\DDDTwo}{0.77}
\newcommand{\StarK}{1.3}
\newcommand{\StarVal}{-5.7}
\newcommand{\StarCallChannel}{-1.9}
\newcommand{\StarRunsSeason}{0.9}
\newcommand{\StarValTwentyThreeOnly}{-22.4}
\newcommand{\StarPreAbs}{1.5}

\begin{document}

\begin{titlepage}
\begin{center}
{\Large\bfseries Who Gains When the Umpire Is a Machine?\\
Evidence from the Automated Ball-Strike System (ABS)\par}
\vspace{2em}
{\large Seung Jin Cho}\\
Department of Economics, Dankook University\\
ORCID: \href{https://orcid.org/0000-0003-1313-0296}{0000-0003-1313-0296}\\[1.5em]
{\large Jun Yeong Lee}\\
Pusan National University\\
ORCID: \href{https://orcid.org/0000-0002-8905-3760}{0000-0002-8905-3760}\\[2em]
\end{center}
\noindent\textbf{Contact information}\par
\medskip
{\setlength{\parindent}{0pt}\singlespacing
Seung Jin Cho\\
Department of Economics, Dankook University\\
152 Jukjeon-ro, Suji-gu, Yongin-si, Gyeonggi-do 16890, Republic of Korea\\
Tel: +82-31-8005-3392\quad Email: \href{mailto:seungjin.cho@dankook.ac.kr}{seungjin.cho@dankook.ac.kr}

\bigskip
Jun Yeong Lee (corresponding author)\\
Room 803, Economics and International Commerce Building, Pusan National University\\
2 Busandaehak-ro 63beon-gil, Geumjeong-gu, Busan 46241, Republic of Korea\\
Tel: +82-51-510-7054\quad Email: \href{mailto:junyeonglee@pusan.ac.kr}{junyeonglee@pusan.ac.kr}\par}

\vfill
\noindent\today
\end{titlepage}

\begin{center}
{\large\bfseries Who Gains When the Umpire Is a Machine?\\
Evidence from the Automated Ball-Strike System (ABS)\par}
\end{center}
\section*{Abstract}
\noindent In 2024 the Korea Baseball Organization became the first top-tier league to replace the home-plate umpire's ball--strike judgment with an automated system. Using over a million pitches from 2021 to 2026, we separate the change into the removal of human discretion and the imposition of a new zone geometry. The machine eliminated count-dependent calls, the value of catcher framing and a small advantage for star batters, rents worth only a few runs per player-season. Because the machine's zone scales with height and treats every matchup alike, it redistributed strikes toward tall batters and across matchups; but few pitches are exposed to these changes, and their effects on plate-appearance outcomes are precisely estimated and close to zero. Pitchers re-optimized quickly by throwing higher, with left-handers catching up to an existing trend, whereas we find no evidence that batters adjusted. The results suggest that the incidence of replacing human judgment with a rule depends on its geometry and on exposure.

\medskip
\noindent\textbf{Keywords:} automated ball-strike system; algorithmic decision-making; evaluator bias; rents; adaptation\\
\textbf{JEL classification:} Z22, J24, D83, O33
\clearpage

\section{Introduction}\label{sec:intro}

Algorithms increasingly replace human judgment in decisions about people, from hiring and lending to bail and grading. A large literature asks whether such algorithms decide better than the humans they replace \citep{Kleinberg2018,Hoffman2018}. Less is known about what the replacement does to the people being evaluated. Removing human discretion should eliminate the advantages that discretion created, but a rule also has a shape. The shape determines who gains and who loses, and those affected may or may not change their behavior in response. Answering these questions requires a setting in which every decision is recorded, the rule that replaces human judgment is known, and the behavior of those being evaluated can be observed before and after the change.

Professional baseball offers such a setting. Every pitch is tracked, every ball--strike call (the umpire's ruling on a pitch the batter does not swing at) is recorded, and the call is the outcome of a judgment that, until recently, was made by a human umpire. In 2024 the Korea Baseball Organization (KBO) became the first top-tier professional league to hand that judgment entirely to an automated ball-strike system (ABS). The ABS calls a strike whenever any part of the ball passes through a rectangle defined by each batter's registered height. We study this change using every regular-season pitch thrown in the KBO from 2021 to 2026, more than a million pitches. We combine the pitches with the tracking system's trajectories, the umpires assigned to each game, and the catcher on the field for each pitch.

We organize the analysis around a simple decomposition. Relative to a human umpire, the machine changes calls in two ways. It removes the \emph{discretionary} component of human judgment, that is, the part of the call that responds to context such as the count (the number of balls and strikes so far in the plate appearance), the batter's reputation or the catcher's receiving skill. It also imposes a new \emph{geometry}, the difference between the zone the rule defines and the zone humans actually called. The first component is a set of rents created by the evaluator; the second is a redistribution chosen by the rule's designers. To a first approximation, the economic incidence of either component equals the number of calls it changes per plate appearance (a batter's turn at bat) times the value of a ball relative to a strike. Finally, those affected should adapt only when the gain from adapting exceeds its cost.

Three sets of findings emerge. First, the machine removed the rents created by human discretion. On border pitches, those close to the edge of the zone, human umpires were \CountAheadPre{} percentage points more likely to call a strike when the batter was ahead in the count (2-0, 3-0 or 3-1) and \CountTwoPreAbs{} points less likely with two strikes; neither adjustment is detectable under the ABS. Catcher framing, the skill of receiving pitches so that umpires call them strikes, was persistent under human umpires, with split-half reliabilities between \FrameRelMin{} and \FrameRelMax, and it became indistinguishable from noise under the machine. Human umpires also gave star batters slightly fewer called strikes on border pitches, although the size of this effect depends on how finely location is held fixed. Valued at what a strike costs the batting team in expected runs, these rents amounted to a few runs per catcher-season and about one run per star batter-season.

Second, the geometry of the rule redistributed strikes along two margins, but the redistribution had little effect on outcomes. Because the human zone was anchored in absolute height while the machine's zone scales with registered height, the ABS moved strikes toward taller batters at the top of the zone: relative to the human zone, each centimeter of height added \FSPooled{} percentage points to the share of a batter's high pitches that were called strikes. Because human umpires treated pitcher--batter matchups asymmetrically, the machine also moved strikes between handedness matchups at the edges. Few pitches are exposed to these changes, however. The extra strikes amount to at most one in every 23 plate appearances, and ten centimeters of height change the probability that a plate appearance reaches two strikes or three balls, one call away from a strikeout or a walk, by less than two percentage points.

Third, pitchers adapted and batters did not. Using the change in each location's value under the machine as a measure of treatment intensity, we find that pitchers reallocated their pitches toward locations the ABS favored, quickly and almost entirely through pitch height. The acceleration at the time of the ABS comes from left-handed pitchers. Right-handed and side-arm pitchers (those who throw with a low, sideways arm angle) had already been raising their pitches before 2024, whereas left-handed pitchers had not, and they caught up within a season. For batters, by contrast, we detect no change in swing decisions, either in response to the new geometry or to the end of count-dependent calls. This asymmetry is consistent with a simple cost of re-optimization. A pitcher's plan is set in advance and applies to every batter, whereas a batter decides in real time, and the stakes for any one batter are small.

The paper contributes to four literatures. A body of work in economics and management documents that human evaluators in sport are influenced by race \citep{PriceWolfers2010,Parsons2011,Tainsky2015,HamrickRasp2015}, status \citep{KimKing2014,Mills2014}, nationality \citep{Zitzewitz2006}, the sequence of past decisions \citep{Chen2016}, the count \citep{MacMahonStarkes2008}, and the home crowd \citep{SutterKocher2004,Garicano2005,Dohmen2008,Hsu2024}, and that the quality of their judgment varies with working conditions \citep{Fesselmeyer2021}. A related set of studies shows that these biases respond to the evaluator's environment: favoritism fell when referees became full-time professionals \citep{RickmanWitt2008} and when referees became aware of their bias \citep{PopePriceWolfers2018}, and pitch-tracking technology changed how umpires called the zone \citep{Mills2017,Hunter2018,Bradbury2019}. These studies observe human evaluators under different incentives and degrees of scrutiny. We observe the complete replacement of the evaluator, which allows us to measure the rents that discretion created and to test whether they vanish. Our results also relate to the economics of algorithmic decision-making \citep{Kleinberg2018,Hoffman2018}. That literature compares the quality of human and machine decisions. We show that the distributional consequences of replacing humans with a rule are governed by the rule's geometry and by exposure, and that those evaluated by the rule may not adapt to it. Third, we add to the evidence on how professional agents adjust their strategies when payoffs change. Some studies find that professionals play close to optimally \citep{WalkerWooders2001,Chiappori2002,PalaciosHuerta2003,Abramitzky2012} and change their strategies as theory predicts when the rules change \citep{Banerjee2007}, while others document persistent departures from optimal play \citep{Romer2006,KovashLevitt2009}. That pitchers adapted to the ABS while batters did not suggests that which pattern prevails depends on how much an agent stands to gain from re-optimizing. Finally, we build on recent studies of the ABS itself. \citet{LeeHanKo2025} describe how the zone changed in the first half of 2024, and \citet{LeeKo2026} show that count-dependent calls disappeared, a result we replicate. \citet{Song2026} attribute a decline in star batters' performance after the ABS to the loss of umpire favoritism. We find evidence of this favoritism at the level of individual calls, but its value appears small, and much of the decline in outcomes appears to reflect the choice of a single pre-ABS season as the baseline.

The rest of the paper proceeds as follows. Section~\ref{sec:framework} develops the conceptual framework and the hypotheses. Section~\ref{sec:data} describes the institutional setting and the data, and Section~\ref{sec:strategy} the empirical strategy. Section~\ref{sec:results} presents the results and their robustness, and Section~\ref{sec:conclusion} concludes.

\section{Conceptual Framework and Hypotheses}\label{sec:framework}

\subsection{Setup}

Consider a pitch that the batter does not swing at, a \emph{taken} pitch. It crosses the middle plane of home plate at location $\ell$ (horizontal position and height), is thrown to a batter of registered height $h$, and arrives in a context $\theta$. The context comprises the count, the batter's reputation, the catcher, and the home-plate umpire. The evaluator calls a strike with probability $S(\ell,h,\theta)$.

Under a human umpire, $S^{H}(\ell,h,\theta)$ is a smooth function of location. It may also depend on the context, through count adjustments, deference to reputation, or the catcher's receiving skill, and it need not scale with the batter's height. Under the ABS, the call is the indicator
\begin{equation}
S^{M}(\ell,h)=\mathbf{1}\{\ell\in Z(h)\}.
\end{equation}
Here $Z(h)$ is the zone defined by the league's rules \citep{KBO2026}. Its vertical bounds are fixed fractions of registered height, and its width is the plate (43.18\,cm) plus 2\,cm on each side; Section~\ref{sec:data} gives the details. By construction, $\partial S^{M}/\partial\theta=0$. The switch from $S^{H}$ to $S^{M}$ in 2024 is the shock we study.

Let $v_{s}(\theta)$ and $v_{b}(\theta)$ be the batter's expected plate-appearance value after a called strike and after a called ball in count $\theta$, and let $\Delta v(\theta)=v_{b}(\theta)-v_{s}(\theta)>0$. We estimate $v$ from 2022--2023 plate appearances, measuring the value of a plate appearance by weighted on-base average (wOBA), which weights each outcome (walk, single, home run and so on) by its average contribution to run scoring and gives outs a weight of zero. We call the weight of a plate appearance's final outcome its \emph{plate-appearance value}; a player's wOBA is the average of these values. On average, one strike is worth \RunsPerStrike{} runs.

\subsection{Two components of the shock}

Let $\bar S^{H}(\ell,h)$ be the human call probability averaged over contexts. The change in the call probability for a given pitch decomposes into
\begin{equation}\label{eq:decomp}
S^{M}(\ell,h)-S^{H}(\ell,h,\theta)
=\underbrace{\big[S^{M}(\ell,h)-\bar S^{H}(\ell,h)\big]}_{\text{geometric}}
-\underbrace{\big[S^{H}(\ell,h,\theta)-\bar S^{H}(\ell,h)\big]}_{\text{discretionary}}.
\end{equation}
The \emph{geometric} component reflects the difference between the rule's zone and the human zone. It varies only with location, height and handedness, and it is a redistribution chosen by the rule's designers. The \emph{discretionary} component is the part of human judgment that responds to context, and the machine sets it to zero. It is a set of rents created by the human evaluator: deference to stars, adjustments to the count, and the value of a catcher's ability to influence calls.

\subsection{Hypotheses}

Our first hypothesis, H1, is that the machine removes discretion. Holding location fixed, called-strike probabilities under the ABS do not depend on the count, the batter's reputation or the catcher. The pre-ABS size of each context effect measures a rent that the machine removes. Catcher framing is a skill whose value comes from the evaluator's errors, so it should be a persistent catcher attribute before the ABS and indistinguishable from noise after it.

The second hypothesis, H2, is that the geometric incidence of the change follows the rule's design. Because $Z(h)$ scales with registered height and is symmetric across handedness, the ABS shifts strikes toward tall batters at the top of the zone. It also removes any handedness asymmetry in the human zone.

The third hypothesis, H3, is that the economic incidence of the change equals exposure times price. To first order, the change in a batter's expected plate-appearance value is
\begin{equation}\label{eq:incidence}
\Delta V\;\approx\;-\sum_{\text{taken pitches}}\Delta S(\ell,h,\theta)\,\Delta v(\theta)\;=\;-E\times\overline{\Delta v},
\end{equation}
where $E$ is the number of calls per plate appearance that the shock flips. Large changes in call probabilities therefore translate into small changes in outcomes when the affected pitches are rare. Where we observe an outcome change for a group, the call channel in \eqref{eq:incidence} bounds the part of that change attributable to the ABS.

The fourth hypothesis, H4, concerns pitchers, who should reallocate their pitches toward locations whose value rose, with partial adjustment. We divide the plate into five vertical bands (the three thirds of the ABS zone and the areas above and below it) and five horizontal bands (the three thirds of its width and the areas beyond each side). A \emph{cell} $c$ is a pitch type (fastball, slider, curveball and so on) in one of the 25 resulting areas, and $s_{t}(c)$ is the share of that pitch type's pitches thrown to the area in season $t$. The ABS changes the value of cell $c$ by its \emph{mechanical shock} $G(c)$, the change in the cell's value to the pitcher, in wOBA points per 1,000 pitches, when its 2022--2023 called pitches are re-judged by the ABS rule with behavior held fixed. We model usage as partial adjustment toward a target, the reduced form implied by convex adjustment costs \citep{Lucas1967,Gould1968}:
\begin{equation}\label{eq:partial}
\Delta s_{t}(c)=\lambda\,\big[s^{*}_{t}(c)-s_{t-1}(c)\big],\qquad s^{*}_{2024}(c)-s^{*}_{2023}(c)=\kappa\,G(c).
\end{equation}
Three implications follow. First, the 2024 change in usage responds to $G$ over and above any pre-existing trend: if $\beta_{t}$ denotes the slope of the usage change on $G$ in season $t$, the acceleration $A=\beta_{2024}-\beta_{2023}$ is positive. Second, the response fades once the new target is reached, so $\beta_{2025}$ and $\beta_{2026}$ are close to zero; a small negative value would indicate overshooting followed by correction. Third, groups that were already moving toward the new target before 2024 show little acceleration, whereas groups that had not moved catch up. These implications require only that usage move in proportion to its distance from the target, not that it move slowly; they hold for any adjustment speed $\lambda$, including one close to one.

The fifth hypothesis, H5, is that agents re-optimize only when the gain exceeds the cost. Even professionals can leave sizable gains unexploited \citep{Romer2006}, which suggests that changing a strategy is costly. Suppose that changing a decision rule carries a fixed cognitive or practice cost in addition to the convex cost above, as in models that combine the two \citep{CooperHaltiwanger2006}. An agent then adjusts only when the expected gain exceeds the fixed cost, so that adjustment is lumpy \citep{CaballeroEngel1999}, and the convex cost governs how quickly an agent that does adjust approaches its target. The two sides of the plate face different gains. A pitcher's location plan is set in advance and applies to every batter, so its gain scales with the league-wide movement of the zone. A batter's swing decision is made in a fraction of a second, and its gain scales with the batter's own exposure $E$, which H3 implies is small. We therefore expect pitchers to adapt and batters not to, even after the count stops affecting calls.
\section{Institutional Setting and Data}\label{sec:data}

\subsection{The ABS in the KBO}

Since the start of the 2024 season the KBO has used the ABS in every regular-season game. A tracking system measures every pitch, and a strike-determination system relays the call to the umpire through an earpiece. Calls cannot be challenged.

The rule defines the zone relative to each batter's registered height. In 2024 the bottom and top of the zone were 27.64\% and 56.35\% of height. From 2025 both bounds moved down by 0.6 percentage points, to 27.04\% and 55.75\%, and they were unchanged in 2026. Horizontally, the zone is the plate widened by 2\,cm on each side, 47.18\,cm in total. According to the league's rules, the horizontal bounds are checked at the middle plane of the plate, and a strike requires the pitch to meet the vertical bounds at both the middle plane and the back plane \citep{KBO2026}. The published rules do not say whether a pitch that only partly overlaps the zone is a strike.

Three other rule changes overlap with our sample. In 2024 the league limited defensive shifts (moving infielders toward the side of the field to which a batter tends to hit) and enlarged the bases; a pitch clock (a time limit between pitches) was piloted in the same year, adopted in 2025 and relaxed in 2026; and in 2022, two years before the ABS, the league instructed umpires to call a wider zone. None of these changes enters the call function $S$, so tests based on calls are unaffected by them. Tests based on behavior and outcomes rely on contrasts within the league, by height, handedness, exposure and pitching arm, so that league-wide shocks are absorbed by season fixed effects.

\subsection{Pitch-level data}

We collect every regular-season pitch from 2021 to 2026 from the play-by-play feed of Naver Sports, the sports service of the Korean web portal Naver, which publishes the league's tracking data \citep{Naver2026}. For each pitch we observe the call or batted-ball outcome, the pitch type and velocity, and the nine parameters of the fitted trajectory. We also observe the batter's registered height and batting side (left or right), the pitcher's throwing arm, and the result of the plate appearance. We add the four umpires listed for each game and the catcher on the field for each pitch, which we reconstruct from the starting lineup and the substitution log.

We restrict the sample to April--September, so that each season covers the same months, and drop the All-Star Game. The final sample contains \NPitches{} pitches, of which \NCalled{} were called, and \NPA{} plate appearances. Table~\ref{tab:descriptives} summarizes them by season. Among other statistics, it reports the share of \emph{high} pitches, those in the upper third of the ABS zone or above (the top two of the five vertical bands defined in Section~\ref{sec:framework}).

\begin{table}[H]
\centering
\caption{\label{tab:descriptives}Summary statistics by season}
\centering
\fontsize{10}{12}\selectfont
\begin{threeparttable}
\begin{tabular}[t]{lcccccc}
\toprule
\multicolumn{1}{c}{ } & \multicolumn{3}{c}{Human umpire} & \multicolumn{3}{c}{ABS} \\
\cmidrule(l{3pt}r{3pt}){2-4} \cmidrule(l{3pt}r{3pt}){5-7}
  & 2021 & 2022 & 2023 & 2024 & 2025 & 2026\\
\midrule
Games & 596 & 687 & 648 & 680 & 674 & 665\\
Pitches & 182,137 & 206,151 & 197,826 & 210,403 & 203,232 & 204,006\\
Called pitches & 101,737 & 110,747 & 107,953 & 113,949 & 110,771 & 112,044\\
Plate appearances & 46,287 & 53,252 & 50,698 & 54,074 & 52,261 & 52,387\\
Called-strike rate (\%) & 31.4 & 32.2 & 32.3 & 32.4 & 33.2 & 33.0\\
Agreement with ABS rule (\%) & 89.8 & 91.4 & 91.6 & 98.6 & 98.6 & 98.6\\
Strikes called outside the zone (\%) & 5.9 & 6.2 & 5.2 & 0.6 & 0.4 & 0.5\\
Balls called inside the zone (\%) & 18.6 & 13.7 & 14.9 & 3.1 & 3.2 & 3.1\\
High pitches (\%) & 28.1 & 28.9 & 30.2 & 34.6 & 34.5 & 33.7\\
Swing rate (\%) & 43.8 & 45.9 & 44.9 & 45.5 & 45.1 & 44.7\\
Walk rate (\%) & 10.5 & 8.6 & 8.9 & 9.0 & 8.9 & 9.2\\
Strikeout rate (\%) & 18.0 & 18.6 & 17.7 & 18.8 & 19.6 & 19.4\\
\bottomrule
\end{tabular}
\begin{tablenotes}[para]
\item \textit{Notes:} 
\item Columns are regular seasons, April--September; 2021 is used only for pre-trend tests, 2022--2023 is the human-umpire baseline and 2024--2026 are ABS seasons. Agreement compares each call with our reconstruction of the ABS rule (Appendix Table \ref{tab:calibration}), applying the 2024 zone through 2024 and the 2025 zone thereafter, and high pitches are those in the upper third of the zone or above.
\end{tablenotes}
\end{threeparttable}
\end{table}

The seasons play different roles. The two seasons 2022--2023 form the human-umpire baseline, and every quantity defined relative to the human zone is estimated on them alone. The most important is the human zone itself, $\bar S^{H}(x,z)$, the probability of a called strike at horizontal position $x$ and height above the ground $z$. We estimate it from 2022--2023 calls with a logit model in which this probability is a smooth function of $x$ and $z$ (a generalized additive model with a tensor-product spline). Because $z$ is measured in absolute terms, the estimated zone is free to ignore the batter's height. Treating the human zone as a smooth probability of a strike, rather than a sharp boundary, follows the evidence that umpires' calls behave like a psychophysical judgment that becomes less certain near the edge \citep{Flannagan2024}. The count values $v$ and the mechanical shocks $G$ are also estimated on these two seasons. The seasons 2024--2026 are the ABS seasons. We use 2021 only for pre-trend tests: it precedes the 2022 zone expansion, and the season was interrupted by COVID-19 and resumed without extra innings (the additional innings played when a game is tied after nine).

\subsection{Reconstructing locations and the ABS rule}

We compute the location at which each pitch crosses the plate from its fitted trajectory and express its height as a fraction of the batter's registered height, $z/h$, the unit in which the ABS zone is defined. Because the published rules leave some details open, Appendix Table~\ref{tab:calibration} searches over candidate definitions and compares each with the 2024--2026 ABS calls. Two features matter. Calls match far better when a strike requires only that some part of the ball touch the zone, as \citet{LeeHanKo2025} assume, than when the whole ball must be inside it (98.6\% against 93.0\%), and the 2\,cm widening adds two percentage points. The choice of plane matters little: checking the vertical bounds at the middle plane alone or, as the rules state, at both the middle and the back plane gives 98.6\% and 98.5\%. We use the middle plane alone, which fits marginally better and gives each pitch a single location. This rule reproduces \CalibAgree\% of calls using registered heights, and \AgreeABSHeight\% using the height list the system itself applies, which we back out from the tracking data's zone bounds (Appendix Tables~\ref{tab:replication} and~\ref{tab:heights}). Applied to the human-umpire seasons, the same reconstruction agrees with \AgreeTwentyTwo\% and \AgreeTwentyThree\% of calls in 2022 and 2023; most of the disagreements are balls called inside the zone.

Two measurement issues require attention. The first is a change in the source format: from games on 16 July 2026 the source data report trajectories relative to a reference point 55\,ft from the plate instead of 50\,ft. Ignoring this would inflate plate-crossing heights by 12--19\,cm. After the correction, the replication rate is identical before and after the switch (Appendix Table~\ref{tab:replication}).

The second is the stability of the tracking calibration, since a recalibration of the vertical axis could masquerade as pitchers throwing higher. Within a pitcher and pitch type, the height of the ball 50\,ft from the plate, close to the release point (where the ball leaves the pitcher's hand), changes by at most 1.8\,cm across seasons (Appendix Table~\ref{tab:withinpitcher}). This is small relative to the changes in plate-crossing height that we attribute to behavior.
\section{Empirical Strategy}\label{sec:strategy}

The ABS was introduced in every game at once, so no part of the league remained under human umpires after 2024. Our designs therefore exploit variation in how strongly the change in the call function bears on different pitches, players and locations, and they lean on the fact that the call function itself is untouched by the league's other rule changes. This section describes each design and the assumption under which it identifies the effect of the ABS. Regressions of binary outcomes are linear probability models. Because we test many hypotheses, we also compute, within the family of tests reported in each table, $q$-values adjusted for the false discovery rate with the procedure of \citet{BenjaminiHochberg1995}.

\subsection{Calls}

To locate the change in the call function, we divide called pitches into eight regions defined relative to the 2024 ABS zone and the batter's registered height: the three thirds of the zone; the high and low edges, bands just above and below the zone; the inside and outside edges, bands just beyond the side of the zone nearer to and farther from the batter; and everything farther out. For a called pitch $i$ in region $r$, thrown by pitcher $p$ to batter $b$ in month $m$ of season $t$, we estimate
\begin{equation}\label{eq:region}
\text{Strike}_{i}=\alpha_{r}+\delta_{r}\,\text{ABS}_{t}+\mu_{p}+\mu_{b}+\mu_{m}+\varepsilon_{i},
\end{equation}
where $\text{Strike}_i$ indicates a called strike, $\text{ABS}_t$ indicates the 2024--2026 seasons, and $\mu_p$, $\mu_b$ and $\mu_m$ are pitcher, batter and month fixed effects. The coefficient $\delta_r$ is the change in the probability that a pitch in region $r$ is called a strike after the machine replaced the umpire, holding the pitcher and the batter fixed. Standard errors in this and every other regression of calls are clustered by game, because pitches in the same game share an umpire, a ballpark and conditions. This is a before--after comparison within a region, and $\delta_r$ measures the effect of the ABS under the assumption that, absent the machine, the probability that a pitch in region $r$ thrown by a given pitcher to a given batter is called a strike would have stayed at its 2022--2023 level. Because the limits on defensive shifts, the larger bases and the pitch clock do not enter the call, these other changes cannot violate the assumption. It would fail if human umpires were already changing their calls, which we check by replacing $\text{ABS}_t$ with indicators for each season from 2021 to 2026, with 2023 as the reference.

The ABS rule treats every pitcher--batter matchup identically. To ask whether human umpires did, we estimate equation~\eqref{eq:region} separately for each combination of pitching arm and batting side. Raw call rates could differ across matchups simply because pitches land in different parts of an edge, so we also modify equation~\eqref{eq:region} to compare pitches at the same location. For each side edge (inside and outside) and each period (2022--2023 and 2024--2026), we replace the regressors of equation~\eqref{eq:region} with indicators for three matchups, whose coefficients we denote $\gamma_k$, and with fixed effects $\mu_\ell$ for the fine location bin $\ell$ of the pitch, 3\,cm horizontally by 2\% of the batter's height vertically. The three matchups are right-handed pitchers facing left-handed batters and left-handed pitchers facing right- and left-handed batters, with right-handed pitchers facing right-handed batters as the reference. Each $\gamma_k$ is then the difference in the called-strike probability between matchup $k$ and the reference matchup at the same location, and a Wald $F$-test of $\gamma_{1}=\gamma_{2}=\gamma_{3}=0$ asks whether all four matchups are called alike. A rule that is symmetric across handedness implies that this null holds under the ABS.

To measure how much the change in calls matters for plate appearances, we compute the exposure $E$ in equation~\eqref{eq:incidence} directly. We re-judge every 2022--2023 called pitch with the ABS rule, holding behavior fixed, and count for each plate appearance the strikes the ABS would have added: pitches inside the ABS zone that umpires called balls, minus pitches outside it that umpires called strikes. Averaging this count over the plate appearances of each matchup gives $E$. This is a calculation rather than an estimate.

The discretionary component is identified by difference-in-differences on \emph{border pitches}, the called pitches within 0.18\,ft (5.5\,cm, about one and a half ball widths) of the ABS zone boundary, where context can plausibly sway a human call:
\begin{equation}\label{eq:did}
Y_{i}=\beta_{1}X_{i}+\beta_{2}\,X_{i}\times\text{ABS}_{t}+\mu_{\ell t}+\mu_{p}+\mu_{u}+\varepsilon_{i},
\end{equation}
where $Y_i=\text{Strike}_i$, $X_i$ is a context variable, $\mu_{\ell t}$ are fixed effects for location bins interacted with season, and $\mu_u$ are fixed effects for the home-plate umpire $u$. We consider three contexts. Following the evidence that status affects umpires' calls \citep{KimKing2014}, we define a \emph{star batter} as one whose on-base percentage (the share of plate appearances in which the batter reaches base by a hit, a walk or being hit by a pitch) was in the top 20\% among batters with at least 300 plate appearances in the previous season. The second context is whether the batter plays for the home team. The third is the count, grouped into hitter-ahead counts (2-0, 3-0 and 3-1), two-strike counts, and all other counts as the reference. The specifications for the home team and the count add batter fixed effects $\mu_b$. For star status we report the estimate with and without them; with them, $\beta_2$ is identified from batters whose status changes. The coefficient $\beta_1$ is the human umpires' context effect at a given location, and $\beta_2$ is its change under the ABS. The identifying assumption is that, absent the ABS, the gap in calls between contexts at the same location would have remained constant. We assess it for star status by replacing $\beta_1X_i+\beta_2X_i\times\text{ABS}_t$ with a separate coefficient on $X_i$ for each season, which shows whether the gap was stable before 2024. Equation~\eqref{eq:did} also serves as the template for several designs below, which change the outcome $Y_i$, the regressors and the fixed effects.

To measure catcher framing, we adapt equation~\eqref{eq:did}. For each season separately and on all called pitches, we replace $\beta_1X_i+\beta_2X_i\times\text{ABS}_t$ with a fixed effect $\phi_j$ for the catcher $j$ who received the pitch, and use fixed effects for location bin, pitcher, batter and home-plate umpire. Because catchers on the same team share pitchers, $\phi_j$ is identified only relative to teammates, so we normalize the effects to have a pitch-weighted mean of zero within each team and keep catchers with at least 1,500 called pitches. We measure the persistence of framing in two ways: the split-half reliability, the correlation between $\phi_j$ estimated on alternating games of the same season, and the correlation between a catcher's $\phi_j$ in consecutive seasons. Each correlation is tested against zero with the $t$-test for a Pearson correlation, which has $n-2$ degrees of freedom for $n$ catchers. To value framing in runs, we multiply each catcher's effect by the catcher's number of called pitches and by the run value of a strike, and take the standard deviation across catchers.

\subsection{Pitchers}

For pitchers we cannot compare treated and untreated groups, because every pitcher faced the machine. Instead we use the fact that the ABS changed the value of some locations much more than others. For each cell $c$ defined in Section~\ref{sec:framework} (a pitch type in one of 25 areas of the plate) with at least 150 pitches in 2022--2023, we compute its mechanical shock $G(c)$ from 2022--2023 pitches alone, and estimate for each season
\begin{equation}\label{eq:cell}
s_{t}(c)-s_{t-1}(c)=a_{f}+\beta_{t}\,G(c)+e_{c,t},
\end{equation}
where $s_t(c)$ is the cell's share, in percent, of its pitch type's pitches, $a_{f}$ are fixed effects for the pitch type $f$ of cell $c$, and cells are weighted by their baseline pitches. This is a continuous-treatment difference-in-differences across cells, with $G$ as the treatment intensity.

A positive $\beta_{2024}$ alone does not identify a response to the ABS, because usage may already have been drifting toward cells that the ABS later favored. We therefore focus on the acceleration $A=\beta_{2024}-\beta_{2023}$, which we estimate by regressing the difference between the 2024 and 2023 changes in usage on $G$ in the same way. It identifies the effect under the assumption that, absent the ABS, the drift would have continued at its 2023 pace. Two checks support the assumption. The slope for 2022 provides a placebo, and, for groups whose 2023 slope is zero, the acceleration reduces to the 2024 slope. To ask whether pitchers did more than move their pitches up or down, we also estimate $A$ with fixed effects for the five vertical bands, which compares cells at the same height. We estimate these slopes for all pitchers, separately by pitching arm, and in a balanced panel of pitchers with at least 300 pitches in at least three seasons. Because cells pool the pitches of many pitchers, we compute standard errors from 999 bootstrap replications that resample pitchers and test the slopes with $z$-tests.

To compare the timing of the move toward high pitches across pitching arms, we regress each pitcher's share of high pitches in a season on pitcher fixed effects and season indicators, with 2023 as the reference. We estimate this regression in the balanced panel, separately for left-handed, right-handed and side-arm pitchers, weighting by pitches and clustering standard errors by pitcher. Each season coefficient $\theta_s$ is the within-pitcher change in the high-pitch share between season $s$ and 2023. To ask whether human umpires treated high pitches differently by pitching arm before the ABS, we adapt equation~\eqref{eq:did} once more, estimating it on called pitches in the upper third of the zone and the high edge in 2022--2023. Its regressors are the interactions of indicators for left-handed and side-arm pitchers with an indicator for 2022, with right-handed pitchers as the reference, and its fixed effects are for location bin, season, pitching arm and batter. The coefficient $\lambda_a$ on each interaction is the change in the called-strike probability from 2023 to 2022 for arm $a$ relative to right-handed pitchers at the same location.

\subsection{Batters}

The ABS scales the zone with registered height, so height determines how strongly the geometric shock bears on a batter. We measure the shock with each batter-season's \emph{excess strike rate} on high pitches, $D_{bt}$. For each high pitch within 1\,ft of the center of the plate that batter $b$ took in season $t$, we subtract from the actual call (one for a strike, zero for a ball) the probability that the 2022--2023 human zone would have called it a strike, and $D_{bt}$ is the average of these differences. For batters observed both before and after the ABS, we estimate
\begin{equation}\label{eq:height}
D_{bt}=\alpha_{b}+\tau_{t}+\pi\,(h_{b}-\bar h)\times\text{ABS}_{t}+u_{bt},
\end{equation}
where $h_b$ is registered height, $\bar h=182$\,cm, $\alpha_b$ and $\tau_t$ are batter and season fixed effects, and batter-seasons are weighted by their number of high pitches. This is the first stage. It is a continuous-treatment difference-in-differences with height as the treatment intensity, and the identifying assumption is parallel trends across heights. We examine the assumption by replacing $(h_b-\bar h)\times\text{ABS}_t$ with $(h_b-\bar h)$ interacted with each season from 2021 to 2026, relative to 2023. Taller batters hit more home runs and strike out more (Appendix Table~\ref{tab:balance}), but batter fixed effects absorb such level differences, and the assumption requires only that their trends do not differ. Standard errors in all regressions for batters are clustered by batter, except where noted.

The reduced form replaces $D_{bt}$ in equation~\eqref{eq:height} with the batter's behavior and plate-appearance outcomes, at the level of the pitch or the plate appearance, keeping $\pi\,(h_{b}-\bar h)\times\text{ABS}_{t}$ and batter fixed effects. When the outcome is whether a pitch is high, we add fixed effects for pitcher by season and for count by season. When it is whether the batter swings at a pitch in the area the ABS added to the zone (above the top of the human zone, 3.35\,ft, and inside the ABS zone), we add fixed effects for absolute location bins (3\,cm by 3\,cm) by season, for count by season, and for pitcher. When it is whether a plate appearance reaches two strikes or three balls, we add fixed effects for pitcher by season.

To test whether batters responded to the end of count-dependent calls, we extend equation~\eqref{eq:did} to a triple difference, estimated on border pitches and pitches in the heart of the zone (inside the zone and more than 0.18\,ft from its boundary). The outcome $Y_i$ is an indicator for a swing, and the regressors are $\text{Border}_i\times\text{ABS}_t$ and its interactions with the hitter-ahead and two-strike count groups, with other counts as the reference. The fixed effects are for location bin by season, count group by season, count group by border, pitcher and batter, and standard errors are clustered by game. The coefficient $\psi_g$ on the triple interaction for count group $g$ is the change under the ABS in the difference in swing rates between border and heart pitches in that count group, relative to other counts. Calls in the heart of the zone did not depend on the count even under human umpires, so they absorb any change in swing behavior by count that is unrelated to the calls. As a placebo, we estimate the same regression on 2022--2023 with 2023 treated as if it were the first ABS season.

Finally, we estimate equation~\eqref{eq:did} at the plate-appearance level for star batters over 2022--2026. The outcome $Y_i$ indicates a walk, a strikeout or reaching base, or is the plate-appearance value; $X_i$ is star status; and the fixed effects are for batter and for pitcher by season. We denote its coefficients $\rho_1$ and $\rho_2$, so that $\rho_2$ is the change in star batters' outcomes under the ABS. We compare $\rho_2$ for the plate-appearance value with the change that the call channel in equation~\eqref{eq:incidence} predicts. The prediction multiplies the pre-ABS star coefficient $\beta_1$ from the call regression by the value $\Delta v$ of each border pitch that star batters took in 2022--2023, sums over these pitches, and divides by star batters' plate appearances in the same seasons. Treating the prediction as fixed, we test whether $\rho_2$ equals it with a $t$-test. To express the call channel in runs per season, we multiply it by 550 plate appearances and divide by the wOBA scale of 1.2. Because earlier work uses 2023 alone as the pre-ABS season, we also estimate this regression with that baseline.
\section{Results}\label{sec:results}

We present the results in three steps. We first characterize what the machine changed in the call function, separating its geometric and discretionary components (H1--H2). We then study how pitchers reallocated their pitches (H4), and finally ask whether batters bore the incidence of the change and whether they responded to it (H3, H5).

\subsection{Calls: geometry and discretion}\label{sec:umpires}

Our reconstruction of the ABS rule reproduces \AgreeTwentyFour\% of ABS calls in 2024 and \AgreeTwentySix\% in 2026, against \AgreeTwentyTwo\% and \AgreeTwentyThree\% of human calls in 2022 and 2023 (Table~\ref{tab:descriptives}). The overall called-strike rate barely changed. The zone did not grow but moved (Appendix Figure~\ref{fig:zonemaps}), as \citet{LeeHanKo2025} document for the first half of 2024.

Table~\ref{tab:geometry} asks where on the plate the calls changed and whether the change follows the rule's geometry. Panel A reports $\delta_r$ from equation~\eqref{eq:region} for each of the eight regions. Every $\delta_r$ is significantly different from zero at the 1\% level, and the signs follow the rule. A pitch in the upper third of the zone became \UpperThirdChange{} percentage points more likely to be called a strike, because human umpires had called a seventh of these pitches balls. A pitch in the band just below the zone became $\LowEdgeChange$ points less likely to be a strike, because human umpires had called a third of them strikes. Pitches in the middle of the zone, already called strikes 95\% of the time, changed little. The season-by-season estimates of equation~\eqref{eq:region} in Appendix Table~\ref{tab:regionevent} show that the change at the top of the zone and on the inside edge arrived in 2024, after two stable human-umpire seasons. On the outside edge and below the zone, by contrast, human calls were already moving in the same direction between 2022 and 2023, so the before--after estimates overstate the effect of the machine there; relative to 2023 alone, the change is $-3.2$ points on the outside edge and $-18.5$ points below the zone.

\begin{table}[H]
\centering
\caption{\label{tab:geometry}The geometry of the change in calls}
\centering
\fontsize{9}{11}\selectfont
\begin{threeparttable}
\begin{tabular}[t]{lccccc}
\toprule
\multicolumn{1}{c}{ } & \multicolumn{2}{c}{Strike rate (\%)} & \multicolumn{2}{c}{ABS effect (pp)} & \multicolumn{1}{c}{ } \\
\cmidrule(l{3pt}r{3pt}){2-3} \cmidrule(l{3pt}r{3pt}){4-5}
 & Human & ABS & Change &  & N\\
\midrule
\addlinespace[0.3em]
\multicolumn{6}{l}{\textit{Panel A. Location}}\\
\hspace{1em}Zone, upper third & 86.2 & 100.0 & 13.9$^{***}$ & (0.3) & 16,453\\
\hspace{1em}Zone, middle third & 94.5 & 98.4 & 4.0$^{***}$ & (0.2) & 20,059\\
\hspace{1em}Zone, lower third & 95.4 & 100.0 & 4.7$^{***}$ & (0.2) & 20,305\\
\hspace{1em}High edge & 16.7 & 21.8 & 5.1$^{***}$ & (0.5) & 13,137\\
\hspace{1em}Low edge & 33.1 & 17.8 & -15.2$^{***}$ & (0.5) & 17,560\\
\hspace{1em}Inside edge & 33.0 & 39.7 & 6.7$^{***}$ & (0.8) & 7,422\\
\hspace{1em}Outside edge & 45.6 & 38.5 & -7.0$^{***}$ & (0.6) & 14,541\\
\hspace{1em}Far outside & 0.9 & 0.0 & -0.7$^{***}$ & (0.1) & 109,223\\
\addlinespace[0.3em]
\multicolumn{6}{l}{\textit{Panel B. Horizontal edges by pitcher hand $\times$ batter side}}\\
\hspace{1em}RHP vs RHB, inside edge & 32.1 & 39.2 & 7.3$^{***}$ & (1.4) & 2,389\\
\hspace{1em}RHP vs RHB, outside edge & 50.4 & 38.9 & -11.4$^{***}$ & (1.0) & 5,122\\
\hspace{1em}RHP vs LHB, inside edge & 32.3 & 39.2 & 6.9$^{***}$ & (1.2) & 2,942\\
\hspace{1em}RHP vs LHB, outside edge & 42.3 & 38.3 & -3.9$^{***}$ & (1.0) & 5,491\\
\hspace{1em}LHP vs RHB, inside edge & 43.4 & 40.0 & -3.2$^{*}$ & (1.8) & 1,287\\
\hspace{1em}LHP vs RHB, outside edge & 41.9 & 37.3 & -4.3$^{***}$ & (1.6) & 1,678\\
\hspace{1em}LHP vs LHB, inside edge & 21.5 & 42.2 & 20.7$^{***}$ & (2.1) & 799\\
\hspace{1em}LHP vs LHB, outside edge & 45.1 & 39.2 & -5.8$^{***}$ & (1.5) & 2,235\\
\hspace{1em}Equal rates across matchups, inside edge: $F$ [$p$] & 25.8 [$<$0.001] & 1.4 [0.227] &  &  & \\
\hspace{1em}Equal rates across matchups, outside edge: $F$ [$p$] & 11.7 [$<$0.001] & 3.2 [0.021] &  &  & \\
\addlinespace[0.3em]
\multicolumn{6}{l}{\textit{Panel C. Mechanical exposure: extra strikes per plate appearance}}\\
\hspace{1em}RHP vs RHB &  &  & 0.007 &  & 37,814\\
\hspace{1em}RHP vs LHB &  &  & 0.024 &  & 38,792\\
\hspace{1em}LHP vs RHB &  &  & 0.019 &  & 14,197\\
\hspace{1em}LHP vs LHB &  &  & 0.043 &  & 13,060\\
\bottomrule
\end{tabular}
\begin{tablenotes}[para]
\item \textit{Notes:} 
\item The first two columns are raw called-strike rates under human umpires (2022--2023) and the ABS (2024--2026); Change is $\delta_r$ from equation \eqref{eq:region} (a linear probability model with region, pitcher, batter and month fixed effects), and N counts 2022--2023 called pitches (plate appearances in Panel C); pp denotes percentage points. Panel A covers every region of the plate, defined relative to the 2024 zone and registered height; Panel B re-estimates the model for the side edges by pitcher hand $\times$ batter side (RHP and LHP are right- and left-handed pitchers, RHB and LHB right- and left-handed batters), and its last two rows give Wald $F$-statistics [$p$-values] for $\gamma_1=\gamma_2=\gamma_3=0$ in the modified equation \eqref{eq:region} of Section \ref{sec:strategy}, that is, for equal call rates across matchups at the same location; Panel C re-judges 2022--2023 called pitches with the ABS rule and reports the extra strikes per plate appearance. $^{***}$ $p<0.01$, $^{**}$ $p<0.05$, $^{*}$ $p<0.1$ in two-sided $t$-tests with standard errors clustered by game.
\end{tablenotes}
\end{threeparttable}
\end{table}

\begin{figure}[H]\centering
  \includegraphics[width=0.75\textwidth]{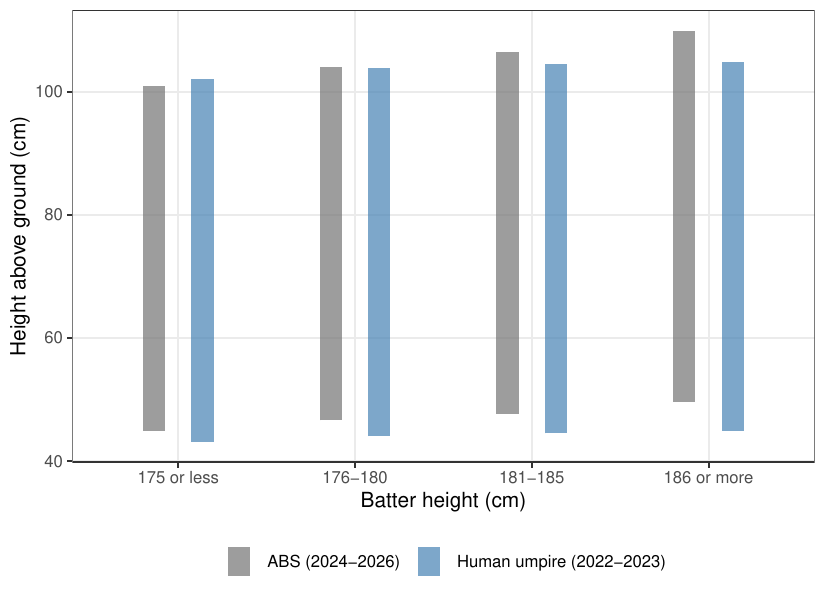}
  \caption{Effective strike-zone height by batter height}\label{fig:zoneheight}
  \begin{minipage}{0.95\textwidth}\footnotesize\textit{Notes:} Called pitches within 0.4 ft of the center of the plate, 2022--2026.
  Bars span the heights at which a logit smooth of the called-strike probability exceeds 0.5.\end{minipage}
\end{figure}

The reason the top of the zone changed so much is that the human zone was anchored in absolute height. Its top sat at about \ZoneTopShortHuman\,cm for the shortest batters and \ZoneTopTallHuman\,cm for the tallest (Figure~\ref{fig:zoneheight}). The ABS scales the zone with height and lifts the top for the tallest group to \ZoneTopTallABS\,cm.

The ABS rule treats every pitcher--batter matchup identically, and Panel B asks whether human umpires did. It reports $\delta_r$ from equation~\eqref{eq:region} for the inside and outside edges, estimated separately for each combination of pitching arm and batting side. Two asymmetries stand out. On the outside edge, umpires were generous when a right-handed pitcher faced a right-handed batter, calling 50.4\% of such pitches strikes against 42--45\% in the other matchups; the ABS cut this rate by 11.4 points. On the inside edge, they were stingy when a left-handed pitcher faced a left-handed batter, calling 21.5\% of such pitches strikes against 32--43\% elsewhere; the ABS raised this rate by 20.7 points. After the ABS, edge strike rates lie between 37\% and 42\% in every matchup.

The last two rows of Panel B compare the matchups at the same location: they report the Wald $F$-statistic for $\gamma_{1}=\gamma_{2}=\gamma_{3}=0$ in the modified equation~\eqref{eq:region}. Under human umpires equality is rejected, with $F=\EqInsideHumanF$ on the inside edge and $F=\EqOutsideHumanF$ on the outside edge (both $p<0.001$). Under the ABS it cannot be rejected on the inside edge ($F=\EqInsideABSF$, $p=\EqInsideABSp$), and on the outside edge the statistic falls to $F=\EqOutsideABSF$ ($p=\EqOutsideABSp$), consistent with small differences in location within bins. These results indicate that the human zone was asymmetric across handedness, whereas the machine's is close to symmetric, as H2 predicts.

Panel C reports the exposure $E$ by matchup. It ranges from \ExpoRHPRHB{} extra strikes per plate appearance for right-handed pitchers facing right-handed batters to \ExpoLHPLHB{} for left-handed pitchers facing left-handed batters. Large changes in call probabilities at the edges therefore amount to at most one extra strike every 23 plate appearances, because few pitches land on the edges and fewer still are taken. This is the first ingredient of H3.

We turn from the geometry of the rule to the discretion it removed. Table~\ref{tab:discretion} reports $\beta_1$ and $\beta_2$ from equation~\eqref{eq:did}, which test H1. Before the ABS, star batters were \StarPreAbs{} percentage points \emph{less} likely than other batters to receive a called strike on the same pitch ($p<0.01$), and the ABS removed most of this advantage ($\beta_2=\StarPost$, $p<0.01$). The same holds with batter fixed effects, comparing seasons in which a batter does and does not hold star status ($p<0.05$). Estimated season by season, the advantage is present in both baseline seasons, at $\StarYrTwentyTwo$ points in 2022 ($p<0.05$) and $\StarYrTwentyThree$ in 2023 ($p<0.01$), and it shrinks toward zero under the ABS, at $\StarYrTwentyFour$, $\StarYrTwentyFive$ and $\StarYrTwentySix$ points in 2024--2026; only the 2024 estimate is significant, and only at the 10\% level. This pattern argues against a pre-existing trend.

There was no home-team bias to remove: neither $\beta_1$ nor $\beta_2$ for the home team is distinguishable from zero (Appendix Table~\ref{tab:home}). The count, by contrast, mattered a great deal. Human umpires expanded the zone in hitter-ahead counts by \CountAheadPre{} points and contracted it with two strikes by $\CountTwoPre$ points (both $p<0.01$), and the estimates of $\beta_2$ are of the opposite sign and similar size (both $p<0.01$), so that the adjustments essentially disappear under the ABS, replicating \citet{LeeKo2026} and consistent with experimental evidence that the count shapes ball--strike judgments \citep{MacMahonStarkes2008}.

\begin{table}[H]
\centering
\caption{\label{tab:discretion}Context-dependent calls: status, home team and count}
\centering
\begin{threeparttable}
\begin{tabular}[t]{lccc}
\toprule
 & Estimate (pp) & SE & N\\
\midrule
\addlinespace[0.3em]
\multicolumn{4}{l}{\textit{Panel A. Batter status}}\\
\hspace{1em}Star batter, pre-ABS & -1.47$^{***}$ & (0.35) & 200,588\\
\hspace{1em}Star batter $\times$ ABS & 1.15$^{***}$ & (0.38) & 200,588\\
\hspace{1em}Star batter $\times$ ABS, batter fixed effects & 1.02$^{**}$ & (0.46) & 200,568\\
\hspace{1em}Star batter in 2022 & -1.15$^{**}$ & (0.51) & 200,588\\
\hspace{1em}Star batter in 2023 & -1.75$^{***}$ & (0.49) & 200,588\\
\hspace{1em}Star batter in 2024 & -0.43$^{*}$ & (0.26) & 200,588\\
\hspace{1em}Star batter in 2025 & -0.35 & (0.25) & 200,588\\
\hspace{1em}Star batter in 2026 & -0.17 & (0.22) & 200,588\\
\addlinespace[0.3em]
\multicolumn{4}{l}{\textit{Panel B. Count}}\\
\hspace{1em}Hitter-ahead count, pre-ABS & 6.14$^{***}$ & (0.45) & 200,568\\
\hspace{1em}Hitter-ahead count $\times$ ABS & -5.47$^{***}$ & (0.48) & 200,568\\
\hspace{1em}Two-strike count, pre-ABS & -6.74$^{***}$ & (0.29) & 200,568\\
\hspace{1em}Two-strike count $\times$ ABS & 6.90$^{***}$ & (0.31) & 200,568\\
\bottomrule
\end{tabular}
\begin{tablenotes}[para]
\item \textit{Notes:} 
\item Columns report the coefficient in percentage points, its standard error and the number of border pitches (within 0.18 ft of the zone boundary), from linear probability models of a called strike with location-bin $\times$ season, pitcher and home-plate-umpire fixed effects. Panel A compares star batters (top 20\% of on-base percentage among batters with 300+ plate appearances in the previous season) with other batters before and after the ABS, adding batter fixed effects in the third row and estimating the star coefficient season by season below; Panel B compares each count group with neutral counts, with batter fixed effects. $^{***}$ $p<0.01$, $^{**}$ $p<0.05$, $^{*}$ $p<0.1$ in two-sided $t$-tests with standard errors clustered by game.
\end{tablenotes}
\end{threeparttable}
\end{table}

Catcher framing is a skill whose value derives from the evaluator's errors, and a literature in sports analytics measures it with catcher effects on the called-strike probability \citep{DeshpandeWyner2017}. H1 therefore predicts that it should be a persistent attribute of catchers under human umpires and noise under the machine. Table~\ref{tab:framing} reports the dispersion and the reliability of the catcher effects $\phi_j$ defined in Section~\ref{sec:strategy}, and the estimates support both parts. Within team, the dispersion of catcher effects falls from \FrameSDPre{} to \FrameSDPost{} percentage points. In every baseline season, estimates from alternating halves of the season's games are strongly correlated (\FrameRelMin--\FrameRelMax, all $p<0.01$ with 20--22 catchers), and a catcher's estimate predicts the next season's ($p<0.05$). In none of the ABS seasons is either correlation distinguishable from zero, even at the 10\% level.

\begin{table}[H]
\centering
\caption{\label{tab:framing}Catcher framing before and after the ABS}
\centering
\begin{threeparttable}
\begin{tabular}[t]{lccccc}
\toprule
\multicolumn{2}{c}{ } & \multicolumn{1}{c}{Dispersion} & \multicolumn{2}{c}{Reliability} & \multicolumn{1}{c}{Value} \\
\cmidrule(l{3pt}r{3pt}){3-3} \cmidrule(l{3pt}r{3pt}){4-5} \cmidrule(l{3pt}r{3pt}){6-6}
Season & Catchers & SD (pp) & Split-half r & Prior-season r & SD (runs)\\
\midrule
2021 & 20 & 1.06 & 0.81$^{***}$ &  & 4.9\\
2022 & 22 & 0.94 & 0.65$^{***}$ & 0.56$^{**}$ & 4.1\\
2023 & 20 & 0.78 & 0.60$^{***}$ & 0.67$^{***}$ & 4.6\\
2024 & 21 & 0.09 & -0.17 & 0.02 & 0.5\\
2025 & 23 & 0.19 & 0.28 & 0.40 & 0.9\\
2026 & 22 & 0.12 & 0.06 & 0.14 & 0.6\\
\bottomrule
\end{tabular}
\begin{tablenotes}[para]
\item \textit{Notes:} 
\item Columns report the number of catchers with 1,500+ called pitches, the standard deviation (SD) of their within-team effects on the called-strike probability in percentage points (pp), the correlation between estimates from alternating halves of the season's games, the correlation with the same catcher's estimate in the previous season, and the standard deviation in runs (one strike = 0.135 runs). Catcher effects come from season-by-season linear probability models with location-bin, pitcher, batter and umpire fixed effects. $^{***}$ $p<0.01$, $^{**}$ $p<0.05$, $^{*}$ $p<0.1$ in two-sided $t$-tests of zero correlation.
\end{tablenotes}
\end{threeparttable}
\end{table}

Appendix Figure~\ref{fig:framing} shows the collapse. The rents involved were real but small. Valued at the run value of a strike in the KBO (\RunsPerStrike{} runs), one standard deviation of framing was worth about \FrameRunsPre{} runs per catcher-season before the ABS, against \FrameRunsPost{} after, and the star advantage in calls was worth about \StarRunsSeason{} runs per star batter-season. Both effectively disappeared under the ABS, and both are small relative to a team's run total.
\subsection{Pitchers: reallocation and catch-up}\label{sec:pitchers}

\begin{figure}[H]\centering
  \includegraphics[width=\textwidth]{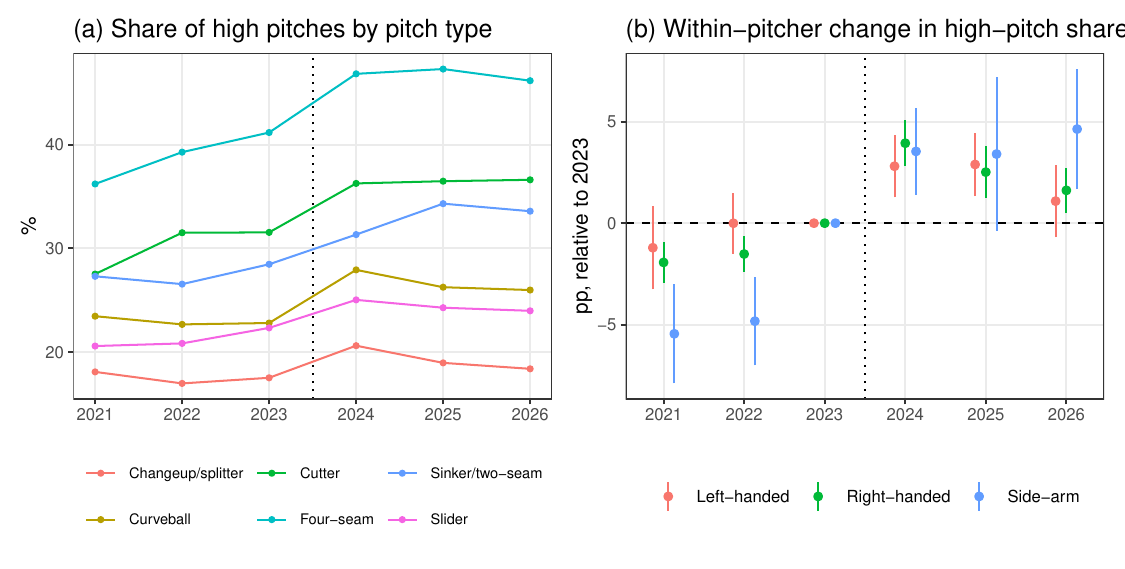}
  \caption{Pitch height before and after the ABS}\label{fig:pitchheight}
  \begin{minipage}{0.95\textwidth}\footnotesize\textit{Notes:} Panel (a): share of all pitches in the upper third of the ABS zone or above,
  by pitch type. Panel (b): season coefficients (2023 = 0) from regressions of a pitcher's high-pitch share on season indicators with
  pitcher fixed effects, balanced panel, by pitching arm; 95\% confidence intervals from standard errors clustered by pitcher; pp denotes percentage points.\end{minipage}
\end{figure}

Pitchers moved their pitches up: the share of pitches in the upper third of the zone or above rose for every pitch type in 2024 (Figure~\ref{fig:pitchheight}a). Table~\ref{tab:reallocation} asks whether this movement was a response to the ABS rather than the continuation of a trend. It reports the slopes $\beta_t$ and the acceleration $A$ from equation~\eqref{eq:cell}; a positive $G$ means that the machine made the cell more valuable to the pitcher.

For all pitchers the response slope is zero in 2022 ($\beta_{2022}=\BetaAllTwentyTwo$, not significant), which serves as a placebo: the ABS shock does not predict changes before the ABS was announced. The slope is positive in 2023 (\BetaAllTwentyThree, $p<0.01$), rises in 2024 (\BetaAllTwentyFour, $p<0.01$), is close to zero in 2025, and turns slightly negative in 2026 ($-0.011$, significant only at the 10\% level), a partial unwinding of the 2024 movement. The acceleration $A=\AccelAll$ is positive but significant only at the 10\% level, and with vertical-band fixed effects it is no longer distinguishable from zero. Pitchers moved their pitches up but did not otherwise re-sort them by pitch type or side of the plate. The adjustment was quick and coarse, consistent with partial adjustment toward a new target that was reached within one season (H4).

\begin{table}[H]
\centering
\caption{\label{tab:reallocation}Pitchers' reallocation toward locations favored by the ABS}
\centering
\fontsize{9}{11}\selectfont
\begin{threeparttable}
\begin{tabular}[t]{lcccccc}
\toprule
\multicolumn{1}{c}{ } & \multicolumn{5}{c}{Response slope $\beta_t$} & \multicolumn{1}{c}{Acceleration} \\
\cmidrule(l{3pt}r{3pt}){2-6} \cmidrule(l{3pt}r{3pt}){7-7}
 & 2022 & 2023 & 2024 & 2025 & 2026 & $A$\\
\midrule
All pitchers & 0.001 & 0.030$^{***}$ & 0.047$^{***}$ & -0.005 & -0.011$^{*}$ & 0.018$^{*}$\\
 & (0.006) & (0.006) & (0.006) & (0.005) & (0.006) & (0.010)\\
Left-handed & -0.007 & 0.003 & 0.060$^{***}$ & 0.002 & -0.021$^{*}$ & 0.058$^{***}$\\
 & (0.011) & (0.009) & (0.011) & (0.010) & (0.011) & (0.015)\\
Right-handed & 0.001 & 0.034$^{***}$ & 0.039$^{***}$ & -0.009 & -0.009 & 0.005\\
 & (0.007) & (0.008) & (0.007) & (0.006) & (0.007) & (0.012)\\
Side-arm & 0.020 & 0.052$^{***}$ & 0.045$^{***}$ & 0.030 & -0.015 & -0.007\\
 & (0.019) & (0.020) & (0.017) & (0.020) & (0.032) & (0.031)\\
Left-handed, balanced & 0.009 & 0.001 & 0.053$^{***}$ & 0.010 & -0.031$^{**}$ & 0.052$^{***}$\\
 & (0.010) & (0.010) & (0.011) & (0.011) & (0.015) & (0.018)\\
Right-handed, balanced & 0.003 & 0.031$^{***}$ & 0.040$^{***}$ & -0.015$^{**}$ & -0.005 & 0.009\\
 & (0.007) & (0.007) & (0.007) & (0.006) & (0.008) & (0.011)\\
All, with vertical-band fixed effects &  &  &  &  &  & -0.012\\
 &  &  &  &  &  & (0.009)\\
\bottomrule
\end{tabular}
\begin{tablenotes}[para]
\item \textit{Notes:} 
\item Each season column reports the slope $\beta_t$ from regressing the change in a cell's usage share between $t-1$ and $t$ on the cell's mechanical shock $G$, with pitch-type fixed effects and baseline-pitch weights, and the last column is the acceleration $A=\beta_{2024}-\beta_{2023}$. Cells are pitch type $\times$ five vertical $\times$ five horizontal bands with 150+ baseline pitches; rows split pitchers by arm, and the balanced panel keeps pitchers with 300+ pitches in at least three seasons. $^{***}$ $p<0.01$, $^{**}$ $p<0.05$, $^{*}$ $p<0.1$ in two-sided $z$-tests with standard errors from 999 bootstrap replications that resample pitchers.
\end{tablenotes}
\end{threeparttable}
\end{table}

The pooled estimate masks sharp heterogeneity by pitching arm. Left-handed pitchers show no response before 2024, a large response in 2024 ($\beta_{2024}=\BetaLHPTwentyFour$, $p<0.01$), none in 2025, and a small reversal in 2026 ($\beta_{2026}=-0.021$, significant only at the 10\% level). Their acceleration is $A=\AccelLHP$, and \AccelLHPbal{} in a balanced panel of the same pitchers (both $p<0.01$). Right-handed pitchers who throw over-hand (with a conventional high arm angle), in contrast, were already responding in 2023 ($\beta_{2023}=\BetaRHPTwentyThree$) and do not accelerate ($A=\AccelRHP$, not significant).

Appendix Table~\ref{tab:lhp} examines why. Its timing rows report the within-pitcher season coefficients $\theta_s$; its last row reports $\lambda_a$, the difference by pitching arm in how human umpires called high pitches; and the rows on the size of the shock, the mix of batters and the type of pitcher are descriptive comparisons. Left-handed pitchers had somewhat more to gain from the ABS, but we find no evidence that they faced a different mix of batters, threw differently, or changed in composition.

What distinguishes them is timing. The within-pitcher coefficients $\theta_s$ (Figure~\ref{fig:pitchheight}b) show that right-handed and side-arm pitchers were already raising their pitches between 2021 and 2023, whereas left-handed pitchers were flat until the ABS arrived. The 2021 and 2022 coefficients relative to 2023 are negative and significant at the 1\% level for right-handed and side-arm pitchers, and indistinguishable from zero for left-handed pitchers. By 2025, all three groups had raised their high-pitch share by similar amounts relative to 2023 (2.5--3.4 points; Appendix Table~\ref{tab:lhp}). For side-arm pitchers, the earlier move coincides with a change in how human umpires called their high pitches: relative to right-handed pitchers, their high pitches were less likely to be called strikes in 2022 than in 2023 ($\lambda_a$ significant at the 1\% level). For right-handed pitchers there is no such coincidence, and we treat their earlier move as a secular trend.

The evidence is consistent with the third implication of H4: the ABS appears to have accelerated reallocation mainly for the group that had not already moved toward the new target. Appendix Figure~\ref{fig:shockresponse} illustrates the contrast. The relationship between the shock and the change in usage is flat for left-handed pitchers before the ABS and steep after it, but similar in both periods for right-handed pitchers.
\subsection{Batters: incidence without adaptation}\label{sec:batters}

Panel A of Table~\ref{tab:batters} reports $\pi$ from the first stage, equation~\eqref{eq:height}, and its season-by-season version. Each centimeter of height raises a batter's excess strike rate on high pitches by \FSPooled{} percentage points after the ABS ($p<0.01$; first-stage Wald $F=\FSF$). The season-by-season estimates jump in 2024 and stay at that level. Before the ABS they are an order of magnitude smaller and move in opposite directions in 2021 and 2022, around the league's 2022 zone expansion. Height therefore captures a large and stable change in the calls a batter receives.

\begin{table}[H]
\centering
\caption{\label{tab:batters}Batters: incidence by height, response to the end of count-dependent calls, and star batters' outcomes}
\centering
\fontsize{9}{11}\selectfont
\begin{threeparttable}
\begin{tabular}[t]{lccc}
\toprule
 & Estimate & SE & N\\
\midrule
\addlinespace[0.3em]
\multicolumn{4}{l}{\textit{Panel A. First stage: excess high strikes (pp) per cm of height}}\\
\hspace{1em}Height $\times$ ABS, pooled 2022--2026 & 0.75$^{***}$ & (0.11) & 1,026\\
\hspace{1em}Height $\times$ 2021 & -0.16$^{**}$ & (0.06) & 1,183\\
\hspace{1em}Height $\times$ 2022 & 0.18$^{***}$ & (0.06) & 1,183\\
\hspace{1em}Height $\times$ 2024 & 0.78$^{***}$ & (0.12) & 1,183\\
\hspace{1em}Height $\times$ 2025 & 0.90$^{***}$ & (0.10) & 1,183\\
\hspace{1em}Height $\times$ 2026 & 0.83$^{***}$ & (0.11) & 1,183\\
\addlinespace[0.3em]
\multicolumn{4}{l}{\textit{Panel B. Reduced form: effect of 10 cm of height after ABS (pp)}}\\
\hspace{1em}High pitch thrown & 0.30 & (0.24) & 867,342\\
\hspace{1em}Swing at a newly created strike & -1.25 & (2.76) & 16,244\\
\hspace{1em}Plate appearance reaches two strikes & 0.57 & (0.59) & 222,828\\
\hspace{1em}Plate appearance reaches three balls & -0.34 & (0.45) & 222,828\\
\addlinespace[0.3em]
\multicolumn{4}{l}{\textit{Panel C. Swing response to the removal of count-dependent calls (pp)}}\\
\hspace{1em}Two strikes $\times$ border $\times$ ABS & 0.77$^{*}$ & (0.42) & 691,755\\
\hspace{1em}Hitter ahead $\times$ border $\times$ ABS & 0.69 & (0.86) & 691,755\\
\hspace{1em}Two strikes $\times$ border $\times$ 2023 (placebo) & -0.85 & (0.65) & 274,154\\
\hspace{1em}Hitter ahead $\times$ border $\times$ 2023 (placebo) & 0.79 & (1.35) & 274,154\\
\addlinespace[0.3em]
\multicolumn{4}{l}{\textit{Panel D. Star batters' outcomes after ABS (pp; wOBA points per 1,000 plate appearances)}}\\
\hspace{1em}Walk & -0.32 & (0.44) & 262,649\\
\hspace{1em}Strikeout & 1.31$^{**}$ & (0.59) & 262,649\\
\hspace{1em}On base & -0.89 & (0.66) & 262,649\\
\hspace{1em}Plate-appearance value (wOBA points) & -5.68 & (6.77) & 262,649\\
\hspace{1em}Predicted by the call channel (wOBA points) & -1.92 &  & \\
\bottomrule
\end{tabular}
\begin{tablenotes}[para]
\item \textit{Notes:} 
\item Columns report the estimate, its standard error and the number of observations; pp denotes percentage points, and wOBA (weighted on-base average) values each plate-appearance outcome by its average contribution to scoring. Panel A regresses each batter-season's excess called-strike rate on high pitches (relative to the 2022--2023 human zone) on height in cm $\times$ ABS with batter and season fixed effects, with season-by-season estimates relative to 2023; Panel B reports the reduced-form effect of 10 cm of height on pitch- or plate-appearance-level outcomes; Panel C is a triple difference in swing probability for border versus zone-interior pitches by count group, with 2023 as a placebo treatment; Panel D compares star and other batters before and after the ABS, and its last row is the change predicted by the call channel. $^{***}$ $p<0.01$, $^{**}$ $p<0.05$, $^{*}$ $p<0.1$ in two-sided $t$-tests with standard errors clustered by batter (Panels A, B and D) or game (Panel C).
\end{tablenotes}
\end{threeparttable}
\end{table}

Despite this, the reduced-form estimates of $\pi$ in Panel B show no detectable effect of height on what happens in the plate appearance. Ten centimeters of height change the probability of reaching two strikes by \RFTwoStrikeTenCm{} points and of reaching three balls by $\RFThreeBallTenCm$ points. Neither is significantly different from zero, and both 95\% confidence intervals exclude effects larger than about 2 points. Pitchers did not throw more high pitches to tall batters, and tall batters did not swing more at the strikes the ABS created above the human zone. These estimates are reasonably precise, and the absence of an effect is consistent with H3, given that a batter takes fewer than one high pitch every two plate appearances.

The removal of count-dependent calls changed the value of taking a border pitch by about seven percentage points in the probability of a strike. Panel C reports the triple-difference coefficients $\psi_g$, which ask whether batters adjusted their swings in response. Relative to pitches in the heart of the zone, border swings with two strikes rose by \DDDTwo{} points. This estimate is significant only at the 10\% level, and the effect is an order of magnitude smaller than the change in calls. There is no response in hitter-ahead counts, and the placebo estimates are flat. Batters did not re-optimize while pitchers did, which is the asymmetry H5 predicts.

\citet{Song2026} attribute a decline in star batters' walks and on-base percentage after the ABS to the loss of umpire favoritism. Our pitch-level evidence is consistent with such favoritism (Table~\ref{tab:discretion}), but its value appears small. Panel D of Table~\ref{tab:batters} reports $\rho_2$, the change in star batters' plate-appearance outcomes under the ABS. Only the strikeout rate (the share of plate appearances ending in three strikes) changes detectably, rising by \StarK{} points ($p<0.05$). The change in plate-appearance value, $\StarVal$ wOBA points per 1,000 plate appearances, is imprecise. It is not significantly different from the $\StarCallChannel$ points that the call channel predicts ($t\approx-0.6$). With a baseline consisting of 2023 alone, as in earlier work, the decline in plate-appearance value is about four times larger ($\StarValTwentyThreeOnly$ points) and significant at the 1\% level (Appendix Table~\ref{tab:starbaseline}); adding 2022 to the baseline removes most of it.

\subsection{Robustness}\label{sec:robustness}

Appendix Table~\ref{tab:robustness} re-estimates the main coefficients under alternative specifications. For the calls, we narrow or widen the border band to 0.12 or 0.25\,ft, coarsen or refine the location bins to 6 or 1.5\,cm, estimate a logit instead of a linear probability model, define star status by the top 10\% or 30\% of on-base percentage or fix it at its 2021 value, and drop 2026. For pitchers, we require 100 or 300 instead of 150 baseline pitches per cell, and for batters we drop 2026. The removal of count-dependent calls is robust to every change. The left-handed pitchers' acceleration is equally stable when we change the minimum number of baseline pitches per cell, and the first stage and the null reduced form for batters are unchanged without 2026.

The removal of the star advantage is less robust. It survives a narrower or wider border band, finer location bins, a broader definition of star status (the top 30\% of on-base percentage) and the exclusion of 2026. It shrinks and loses significance, however, when location bins are coarsened to 6\,cm, when the model is a logit, and when star status is restricted to the top 10\%, and it is significant only at the 10\% level when status is fixed at its 2021 value. The logit is informative here: once the ABS makes calls a deterministic function of location, many location bins are perfectly predicted after 2024 and drop out of the logit, so it identifies the interaction from a different and smaller set of pitches. We read the evidence as showing that human umpires favored star batters on border pitches, but that the size of the effect depends on how finely location is held fixed.

Appendix Table~\ref{tab:fdr} reports the false-discovery-rate adjusted $q$-values. The removal of the star advantage ($q=0.004$), the removal of count-dependent calls ($q<0.001$) and the acceleration of left-handed pitchers ($q<0.001$) survive the adjustment. The two marginal results for batters do not: the increase in star batters' strikeouts ($q=0.27$) and the response of two-strike swings ($q=0.33$). This reinforces our reading that we cannot detect changes in batters' behavior or outcomes.

\subsection{Summary}

The machine removed three rents created by human judgment: deference to star batters, adjustments to the count, and the value of catcher framing (H1). It also redistributed strikes along two geometric margins chosen by its designers, height and handedness (H2). The redistribution was large in terms of call probabilities but small in terms of plate-appearance outcomes, because few pitches are exposed to it (H3). Pitchers re-optimized quickly and coarsely, and the group that had not already moved caught up (H4). We find no detectable response by batters, either to the new geometry or to the end of count-dependent calls (H5).

\section{Conclusion}\label{sec:conclusion}

This paper studies what happens when an algorithm replaces human judgment in a setting where every decision is recorded and the quality of each decision can be measured. In 2024 the KBO replaced the home-plate umpire's ball--strike judgment with an automated system, and we follow more than a million pitches from 2021 to 2026 to trace the consequences for the evaluator's calls, for the pitchers who throw the pitches, and for the batters who face them. The machine changed the call function in two distinct ways. It removed the discretionary component of human judgment. Before 2024, umpires expanded the zone when the batter was ahead in the count and contracted it with two strikes. Their calls depended on the catcher in a way that was stable within and across seasons, and they gave star batters slightly more benefit of the doubt on border pitches, although the size of this last effect depends on how finely location is held fixed. None of the three patterns is detectable under the ABS. At the same time, the machine imposed a geometry chosen by the rule's designers. Because the human zone was anchored in absolute height and the machine's zone scales with registered height, strikes moved toward taller batters at the top of the zone. Because human umpires treated matchups asymmetrically and the machine does not, strikes moved between right- and left-handed matchups at the edges.

These changes were large in terms of call probabilities but small in terms of outcomes. A pitch in the upper third of the zone became \UpperThirdChange{} percentage points more likely to be called a strike. Yet the extra strikes amount to at most one in every 23 plate appearances for the most exposed matchup, and ten centimeters of height change the probability that a plate appearance reaches two strikes or three balls by less than two percentage points. The value of catcher framing and of the star advantage was real, but it amounted to a few runs per catcher-season and about one run per star batter-season. This suggests that the incidence of a change in an evaluation rule may depend on exposure as much as on the size of the change: a rule can move decisions sharply at the margin while leaving the distribution of outcomes nearly intact, because few cases sit at the margin.

The two sides of the plate responded very differently. Pitchers reallocated their pitches toward the locations that the machine made more valuable, and they did so quickly and coarsely. The response was concentrated in the first season, worked almost entirely through pitch height rather than pitch type or side of the plate, and was partly unwound in the third season. Left-handed pitchers, who had not taken part in the league's earlier drift toward high pitches, account for the acceleration at the time of the ABS and converged to the other groups within a season. Batters, by contrast, show no detectable change in their swing decisions, either in response to the new geometry or to the end of count-dependent calls, even though the value of taking a border pitch changed substantially in some counts. This asymmetry is consistent with a simple cost of re-optimization. A pitcher's plan applies to every batter and is set in advance, so its gain scales with the league-wide shift in the zone, whereas a batter's decision is made in real time and its gain scales with the batter's own small exposure.

The findings bear on the design of automated evaluation in sport and beyond. Leagues that adopt automated calls, fully or through challenge systems such as those used in tennis and, since 2026, in Major League Baseball \citep{Abramitzky2012,LeeHanLeeKo2026}, choose a geometry when they write the rule, and that choice shapes who gains and who loses. Scaling the zone with height, defining the plane at which the ball is judged, and widening the plate are design parameters with distributional consequences, not technical details. More generally, our results suggest that replacing a human evaluator with a rule can eliminate the rents that the evaluator's discretion created, including the value of skills whose purpose was to influence that discretion, but need not by itself make outcomes more equal across agents. The incidence of the new rule is likely to depend on its geometry and on how many cases it touches.

Several limitations qualify these conclusions. We observe one league and three seasons after the change, so we cannot say whether batters eventually adapt or whether the partial reversal in pitchers' behavior in 2026 persists. We do not observe salaries or contracts, so the labor-market consequences of the rents we document remain open. Our measure of star status relies on past on-base percentage, which captures reputation imperfectly. Our pitch locations come from the league's tracking data; we validate their stability within pitchers and against the system's own calls, but we cannot verify the pre-2024 coordinates independently. Finally, our analyses of behavior and outcomes rely on contrasts within the league to absorb the other rule changes of 2024--2026, an assumption that the calls themselves do not need. Comparisons with leagues that kept human umpires, and with Major League Baseball's challenge system, in which umpires still make every call but players can ask the machine to overturn some of them \citep{LeeHanLeeKo2026}, would sharpen these results.

\bibliographystyle{apacite}
\bibliography{references}

@article{LeeHanKo2025,
  author  = {Lee, Kichang and Han, Kyungsik and Ko, JeongGil},
  title   = {Analyzing the Impact of the Automatic Ball Strike System in Professional Baseball through a Case Study on {KBO} League Data},
  journal = {Scientific Reports},
  year    = {2025},
  volume  = {15},
  pages   = {44459},
  doi     = {10.1038/s41598-025-28142-y}
}

@misc{LeeKo2026,
  author       = {Lee, Kichang and Ko, JeongGil},
  title        = {Auditing Contextual Bias in Human Ball-Strike Calls Using {KBO}'s Automated Umpiring Transition},
  year         = {2026},
  howpublished = {arXiv preprint 2609.03786}
}

@article{Song2026,
  author  = {Song, Jimin and Kang, Ji Hyuk and Paulsen, Richard J.},
  title   = {Technology Adoption and Bias in Officiating: Automated Ball-Strike System Implementation in {K}orean Baseball},
  journal = {European Sport Management Quarterly},
  year    = {2026},
  doi     = {10.1080/16184742.2026.2681111},
  note    = {Advance online publication},
  pages = {1--23}
}

@article{Parsons2011,
  author  = {Parsons, Christopher A. and Sulaeman, Johan and Yates, Michael C. and Hamermesh, Daniel S.},
  title   = {Strike Three: Discrimination, Incentives, and Evaluation},
  journal = {American Economic Review},
  year    = {2011},
  volume  = {101},
  number  = {4},
  pages   = {1410--1435},
  doi = {10.1257/aer.101.4.1410}
}

@article{KimKing2014,
  author  = {Kim, Jerry W. and King, Brayden G.},
  title   = {Seeing Stars: Matthew Effects and Status Bias in {M}ajor {L}eague {B}aseball Umpiring},
  journal = {Management Science},
  year    = {2014},
  volume  = {60},
  number  = {11},
  pages   = {2619--2644},
  doi = {10.1287/mnsc.2014.1967}
}

@article{Chen2016,
  author  = {Chen, Daniel L. and Moskowitz, Tobias J. and Shue, Kelly},
  title   = {Decision Making Under the Gambler's Fallacy: Evidence from Asylum Judges, Loan Officers, and Baseball Umpires},
  journal = {Quarterly Journal of Economics},
  year    = {2016},
  volume  = {131},
  number  = {3},
  pages   = {1181--1242},
  doi = {10.1093/qje/qjw017}
}

@article{Mills2017,
  author  = {Mills, Brian M.},
  title   = {Policy Changes in {M}ajor {L}eague {B}aseball: Improved Agent Behavior and Ancillary Productivity Outcomes},
  journal = {Economic Inquiry},
  year    = {2017},
  volume  = {55},
  number  = {2},
  pages   = {1104--1118},
  doi = {10.1111/ecin.12396}
}

@article{Hsu2024,
  author  = {Hsu, Mike},
  title   = {Umpire Home Bias in {M}ajor {L}eague {B}aseball},
  journal = {Journal of Sports Economics},
  year    = {2024},
  doi     = {10.1177/15270025231222631},
  volume = {25},
  number = {4},
  pages = {423--442}
}

@article{WalkerWooders2001,
  author  = {Walker, Mark and Wooders, John},
  title   = {Minimax Play at {W}imbledon},
  journal = {American Economic Review},
  year    = {2001},
  volume  = {91},
  number  = {5},
  pages   = {1521--1538},
  doi = {10.1257/aer.91.5.1521}
}

@techreport{KovashLevitt2009,
  author      = {Kovash, Kenneth and Levitt, Steven D.},
  title       = {Professionals Do Not Play Minimax: Evidence from {M}ajor {L}eague {B}aseball and the {N}ational {F}ootball {L}eague},
  institution = {National Bureau of Economic Research},
  type        = {Working Paper},
  number      = {15347},
  year        = {2009},
  address = {Cambridge, MA}
}

@article{Kleinberg2018,
  author  = {Kleinberg, Jon and Lakkaraju, Himabindu and Leskovec, Jure and Ludwig, Jens and Mullainathan, Sendhil},
  title   = {Human Decisions and Machine Predictions},
  journal = {Quarterly Journal of Economics},
  year    = {2018},
  volume  = {133},
  number  = {1},
  pages   = {237--293},
  doi = {10.1093/qje/qjx032}
}

@article{Hoffman2018,
  author  = {Hoffman, Mitchell and Kahn, Lisa B. and Li, Danielle},
  title   = {Discretion in Hiring},
  journal = {Quarterly Journal of Economics},
  year    = {2018},
  volume  = {133},
  number  = {2},
  pages   = {765--800},
  doi = {10.1093/qje/qjx042}
}

@article{Tainsky2015,
  author  = {Tainsky, Scott and Mills, Brian M. and Winfree, Jason A.},
  title   = {Further Examination of Potential Discrimination Among {MLB} Umpires},
  journal = {Journal of Sports Economics},
  year    = {2015},
  volume  = {16},
  pages   = {353--374},
  number = {4},
  doi = {10.1177/1527002513487740}
}

@article{HamrickRasp2015,
  author  = {Hamrick, Jeff and Rasp, John},
  title   = {The Connection Between Race and Called Strikes and Balls},
  journal = {Journal of Sports Economics},
  year    = {2015},
  volume  = {16},
  pages   = {714--734},
  number = {7},
  doi = {10.1177/1527002513509817}
}

@article{Fesselmeyer2021,
  author  = {Fesselmeyer, Eric},
  title   = {The Impact of Temperature on Labor Quality: Umpire Accuracy in {M}ajor {L}eague {B}aseball},
  journal = {Southern Economic Journal},
  year    = {2021},
  volume  = {88},
  pages   = {545--567},
  number = {2},
  doi = {10.1002/soej.12524}
}

@article{Flannagan2024,
  author  = {Flannagan, Kevin S. and Mills, Brian M. and Goldstone, Robert L.},
  title   = {The Psychophysics of Home Plate Umpire Calls},
  journal = {Scientific Reports},
  year    = {2024},
  volume  = {14},
  pages   = {2735},
  doi     = {10.1038/s41598-024-52402-y}
}

@article{Hunter2018,
  author  = {Hunter, David J.},
  title   = {New Metrics for Evaluating Home Plate Umpire Consistency and Accuracy},
  journal = {Journal of Quantitative Analysis in Sports},
  year    = {2018},
  volume  = {14},
  pages   = {159--172},
  number = {4},
  doi = {10.1515/jqas-2018-0061}
}

@article{MacMahonStarkes2008,
  author  = {MacMahon, Clare and Starkes, Janet L.},
  title   = {Contextual Influences on Baseball Ball-Strike Decisions in Umpires, Players, and Controls},
  journal = {Journal of Sports Sciences},
  year    = {2008},
  volume  = {26},
  pages   = {751--760},
  doi     = {10.1080/02640410701813050}
}

@article{DeshpandeWyner2017,
  author  = {Deshpande, Sameer K. and Wyner, Abraham},
  title   = {A Hierarchical {B}ayesian Model of Pitch Framing},
  journal = {Journal of Quantitative Analysis in Sports},
  year    = {2017},
  volume  = {13},
  pages   = {95--112},
  number = {3},
  doi = {10.1515/jqas-2017-0027}
}

@article{BenjaminiHochberg1995,
  author  = {Benjamini, Yoav and Hochberg, Yosef},
  title   = {Controlling the False Discovery Rate: A Practical and Powerful Approach to Multiple Testing},
  journal = {Journal of the Royal Statistical Society: Series B (Methodological)},
  year    = {1995},
  volume  = {57},
  number  = {1},
  pages   = {289--300},
  doi     = {10.1111/j.2517-6161.1995.tb02031.x}
}

@misc{KBO2026,
  author       = {{Korea Baseball Organization}},
  title        = {2026 {KBO} League Major Regulations and Rules},
  year         = {2026},
  howpublished = {\url{https://www.koreabaseball.com/Kbo/League/GameManage2026.aspx}},
  note         = {In Korean. Accessed 14 October 2026}
}

@misc{Naver2026,
  author       = {{Naver Sports}},
  title        = {{KBO} League Game Schedules, Box Scores and Pitch-by-Pitch Relay, 2021--2026},
  year         = {2026},
  howpublished = {\url{https://m.sports.naver.com/kbaseball}},
  note         = {In Korean. Accessed October 2026}
}

@article{Lucas1967,
  author  = {Lucas, Robert E., Jr.},
  title   = {Adjustment Costs and the Theory of Supply},
  journal = {Journal of Political Economy},
  year    = {1967},
  volume  = {75},
  number  = {4},
  pages   = {321--334},
  doi     = {10.1086/259289}
}

@article{Gould1968,
  author  = {Gould, John P.},
  title   = {Adjustment Costs in the Theory of Investment of the Firm},
  journal = {Review of Economic Studies},
  year    = {1968},
  volume  = {35},
  number  = {1},
  pages   = {47--55},
  doi     = {10.2307/2974406}
}

@article{CaballeroEngel1999,
  author  = {Caballero, Ricardo J. and Engel, Eduardo M. R. A.},
  title   = {Explaining Investment Dynamics in {U.S.} Manufacturing: A Generalized {(S,s)} Approach},
  journal = {Econometrica},
  year    = {1999},
  volume  = {67},
  number  = {4},
  pages   = {783--826},
  doi     = {10.1111/1468-0262.00053}
}

@article{CooperHaltiwanger2006,
  author  = {Cooper, Russell W. and Haltiwanger, John C.},
  title   = {On the Nature of Capital Adjustment Costs},
  journal = {Review of Economic Studies},
  year    = {2006},
  volume  = {73},
  number  = {3},
  pages   = {611--633},
  doi     = {10.1111/j.1467-937X.2006.00389.x}
}

@article{Bradbury2019,
  author  = {Bradbury, John Charles},
  title   = {Monitoring and Employee Shirking: Evidence From {MLB} Umpires},
  journal = {Journal of Sports Economics},
  year    = {2019},
  volume  = {20},
  number  = {6},
  pages   = {850--872},
  doi     = {10.1177/1527002518808350}
}

@article{RickmanWitt2008,
  author  = {Rickman, Neil and Witt, Robert},
  title   = {Favouritism and Financial Incentives: A Natural Experiment},
  journal = {Economica},
  year    = {2008},
  volume  = {75},
  number  = {298},
  pages   = {296--309},
  doi     = {10.1111/j.1468-0335.2007.00605.x}
}

@article{PopePriceWolfers2018,
  author  = {Pope, Devin G. and Price, Joseph and Wolfers, Justin},
  title   = {Awareness Reduces Racial Bias},
  journal = {Management Science},
  year    = {2018},
  volume  = {64},
  number  = {11},
  pages   = {4988--4995},
  doi     = {10.1287/mnsc.2017.2901}
}

@article{Abramitzky2012,
  author  = {Abramitzky, Ran and Einav, Liran and Kolkowitz, Shimon and Mill, Roy},
  title   = {On the Optimality of Line Call Challenges in Professional Tennis},
  journal = {International Economic Review},
  year    = {2012},
  volume  = {53},
  number  = {3},
  pages   = {939--964},
  doi     = {10.1111/j.1468-2354.2012.00706.x}
}

@article{Garicano2005,
  author  = {Garicano, Luis and Palacios-Huerta, Ignacio and Prendergast, Canice},
  title   = {Favoritism Under Social Pressure},
  journal = {Review of Economics and Statistics},
  year    = {2005},
  volume  = {87},
  number  = {2},
  pages   = {208--216},
  doi     = {10.1162/0034653053970267}
}

@article{PriceWolfers2010,
  author  = {Price, Joseph and Wolfers, Justin},
  title   = {Racial Discrimination Among {NBA} Referees},
  journal = {Quarterly Journal of Economics},
  year    = {2010},
  volume  = {125},
  number  = {4},
  pages   = {1859--1887},
  doi     = {10.1162/qjec.2010.125.4.1859}
}

@article{SutterKocher2004,
  author  = {Sutter, Matthias and Kocher, Martin G.},
  title   = {Favoritism of Agents---The Case of Referees' Home Bias},
  journal = {Journal of Economic Psychology},
  year    = {2004},
  volume  = {25},
  number  = {4},
  pages   = {461--469},
  doi     = {10.1016/S0167-4870(03)00013-8}
}

@article{Dohmen2008,
  author  = {Dohmen, Thomas J.},
  title   = {The Influence of Social Forces: Evidence from the Behavior of Football Referees},
  journal = {Economic Inquiry},
  year    = {2008},
  volume  = {46},
  number  = {3},
  pages   = {411--424},
  doi     = {10.1111/j.1465-7295.2007.00112.x}
}

@article{Zitzewitz2006,
  author  = {Zitzewitz, Eric},
  title   = {Nationalism in Winter Sports Judging and Its Lessons for Organizational Decision Making},
  journal = {Journal of Economics \& Management Strategy},
  year    = {2006},
  volume  = {15},
  number  = {1},
  pages   = {67--99},
  doi     = {10.1111/j.1530-9134.2006.00092.x}
}

@article{Mills2014,
  author  = {Mills, Brian M.},
  title   = {Social Pressure at the Plate: Inequality Aversion, Status, and Mere Exposure},
  journal = {Managerial and Decision Economics},
  year    = {2014},
  volume  = {35},
  number  = {6},
  pages   = {387--403},
  doi     = {10.1002/mde.2630}
}

@article{Romer2006,
  author  = {Romer, David},
  title   = {Do Firms Maximize? {E}vidence from Professional Football},
  journal = {Journal of Political Economy},
  year    = {2006},
  volume  = {114},
  number  = {2},
  pages   = {340--365},
  doi     = {10.1086/501171}
}

@article{Chiappori2002,
  author  = {Chiappori, Pierre-Andr{\'e} and Levitt, Steven and Groseclose, Timothy},
  title   = {Testing Mixed-Strategy Equilibria When Players Are Heterogeneous: The Case of Penalty Kicks in Soccer},
  journal = {American Economic Review},
  year    = {2002},
  volume  = {92},
  number  = {4},
  pages   = {1138--1151},
  doi     = {10.1257/00028280260344678}
}

@article{PalaciosHuerta2003,
  author  = {Palacios-Huerta, Ignacio},
  title   = {Professionals Play Minimax},
  journal = {Review of Economic Studies},
  year    = {2003},
  volume  = {70},
  number  = {2},
  pages   = {395--415},
  doi     = {10.1111/1467-937X.00249}
}

@article{Banerjee2007,
  author  = {Banerjee, Anurag N. and Swinnen, Johan F. M. and Weersink, Alfons},
  title   = {Skating on Thin Ice: Rule Changes and Team Strategies in the {NHL}},
  journal = {Canadian Journal of Economics},
  year    = {2007},
  volume  = {40},
  number  = {2},
  pages   = {493--514},
  doi     = {10.1111/j.1540-5982.2007.00418.x}
}

@misc{LeeHanLeeKo2026,
  author       = {Lee, Kichang and Han, Gyeongmin and Lee, Sungmin and Ko, JeongGil},
  title        = {When the Strike Zone Becomes Algorithmic: Umpire Judgment and Player Challenge Decisions under {AI} Review},
  year         = {2026},
  howpublished = {arXiv preprint 2609.25525}
}

\clearpage
\appendix
\renewcommand{\thetable}{A\arabic{table}}\setcounter{table}{0}
\renewcommand{\thefigure}{A\arabic{figure}}\setcounter{figure}{0}
\section{Appendix: Measurement and Robustness}
\begin{table}[H]
\centering
\caption{\label{tab:calibration}Reconstructing the ABS rule from 2024--2026 calls}
\centering
\begin{threeparttable}
\begin{tabular}[t]{lllc}
\toprule
Zone width & Ball radius counts & Judging plane & Agreement (\%)\\
\midrule
Plate + 2 cm & Yes & Middle & 98.61\\
Plate + 2 cm & Yes & Middle and back & 98.49\\
Plate + 2 cm & Yes & Back & 98.00\\
Plate & Yes & Middle & 96.61\\
Plate & Yes & Middle and back & 96.31\\
Plate & Yes & Back & 96.13\\
Plate + 2 cm & Yes & Front & 95.67\\
Plate & Yes & Front & 94.80\\
Plate + 2 cm & No & Middle & 93.00\\
Plate + 2 cm & No & Front & 92.84\\
Plate + 2 cm & No & Back & 92.34\\
Plate + 2 cm & No & Middle and back & 91.29\\
Plate & No & Middle & 90.85\\
Plate & No & Front & 90.68\\
Plate & No & Back & 90.20\\
Plate & No & Middle and back & 89.22\\
\bottomrule
\end{tabular}
\begin{tablenotes}[para]
\item \textit{Notes:} 
\item Each row is a candidate definition of the ABS rule, and the last column is the share of 2024--2026 ABS calls it reproduces using registered heights.
\end{tablenotes}
\end{threeparttable}
\end{table}

\begin{table}[H]
\centering
\caption{\label{tab:withinpitcher}Measurement stability: within pitcher-by-pitch-type season effects}
\centering
\begin{threeparttable}
\begin{tabular}[t]{lccc}
\toprule
  & Height at 50 ft (cm) & Plate-crossing height (cm) & Velocity (km/h)\\
\midrule
2021 & -0.22 & -2.20$^{***}$ & -0.14\\
 & (0.33) & (0.26) & (0.18)\\
2022 & -0.70$^{***}$ & -1.29$^{***}$ & 0.46$^{***}$\\
 & (0.25) & (0.25) & (0.12)\\
2024 & 0.73$^{**}$ & 2.85$^{***}$ & -0.51$^{***}$\\
 & (0.36) & (0.26) & (0.13)\\
2025 & 1.76$^{***}$ & 2.49$^{***}$ & 1.27$^{***}$\\
 & (0.55) & (0.31) & (0.19)\\
2026 & -0.32 & 1.82$^{***}$ & 1.42$^{***}$\\
 & (0.55) & (0.34) & (0.21)\\
\bottomrule
\end{tabular}
\begin{tablenotes}[para]
\item \textit{Notes:} 
\item Columns report season coefficients (2023 omitted) from regressions with pitcher $\times$ pitch-type fixed effects; height at 50 ft approximates release height, so a recalibration of the vertical axis would move both heights equally. $^{***}$ $p<0.01$, $^{**}$ $p<0.05$, $^{*}$ $p<0.1$ in two-sided $t$-tests with standard errors clustered by pitcher.
\end{tablenotes}
\end{threeparttable}
\end{table}

\begin{table}[H]
\centering
\caption{\label{tab:replication}Replication of ABS calls by period}
\centering
\begin{threeparttable}
\begin{tabular}[t]{lccc}
\toprule
Period & Called pitches & Registered height (\%) & ABS height list (\%)\\
\midrule
2024 & 113,949 & 98.59 & 99.16\\
2025 & 110,771 & 98.62 & 99.09\\
2026, from 16 July & 43,121 & 98.58 & 99.09\\
2026, to 15 July & 68,923 & 98.66 & 99.13\\
\bottomrule
\end{tabular}
\begin{tablenotes}[para]
\item \textit{Notes:} 
\item Columns report the number of called pitches and the share of ABS calls reproduced using registered heights and using the system's own height list. From 16 July 2026 the source data's trajectory reference point moves from 50 to 55 ft, which we correct.
\end{tablenotes}
\end{threeparttable}
\end{table}

\begin{table}[H]
\centering
\caption{\label{tab:heights}Registered heights and the heights used by the ABS}
\centering
\begin{threeparttable}
\begin{tabular}[t]{lccccc}
\toprule
Season & Batters & Correlation & Slope & Median gap (cm) & Within 2 cm (\%)\\
\midrule
2024 & 170 & 0.958 & 0.961 & 1.30 & 62.9\\
2025 & 186 & 0.955 & 1.019 & 1.15 & 71.0\\
2026 & 186 & 0.951 & 0.950 & 1.08 & 75.3\\
\bottomrule
\end{tabular}
\begin{tablenotes}[para]
\item \textit{Notes:} 
\item Columns compare registered heights with heights backed out from the tracking system's zone bounds, for batters with 200+ pitches in a season.
\end{tablenotes}
\end{threeparttable}
\end{table}

\begin{table}[H]
\centering
\caption{\label{tab:umpireorder}Which listed umpire explains the calls?}
\centering
\begin{threeparttable}
\begin{tabular}[t]{lcc}
\toprule
Umpire position & Increase in adjusted R$^2$ ($\times 1000$) & SD of umpire effects (pp)\\
\midrule
First listed (home plate) & 1.50 & 2.00\\
Second listed & 0.62 & 1.41\\
Third listed & 0.75 & 1.50\\
Fourth listed & 0.73 & 1.51\\
\bottomrule
\end{tabular}
\begin{tablenotes}[para]
\item \textit{Notes:} 
\item Columns report the increase in adjusted $R^2$ from adding fixed effects for the umpire in each listed position to a model with location-bin $\times$ season fixed effects, and the standard deviation (SD) of those effects in percentage points (pp), for border pitches in 2022--2023.
\end{tablenotes}
\end{threeparttable}
\end{table}

\begin{table}[H]
\centering
\caption{\label{tab:balance}Height and batters' baseline profiles}
\centering
\begin{threeparttable}
\begin{tabular}[t]{lccc}
\toprule
Variable & Coefficient per 10 cm & SE & Correlation\\
\midrule
Home-run rate & 0.0100$^{***}$ & (0.0016) & 0.41\\
Strikeout rate & 0.0409$^{***}$ & (0.0090) & 0.32\\
Walk rate & -0.0007 & (0.0039) & -0.01\\
Pulled ground-ball share & 0.0454$^{***}$ & (0.0120) & 0.27\\
\bottomrule
\end{tabular}
\begin{tablenotes}[para]
\item \textit{Notes:} 
\item Columns report the coefficient on height (per 10 cm) and the correlation with height for baseline characteristics of the batters in Table \ref{tab:batters}, where a pulled ground ball is one hit in the direction of the batter's swing (toward left field for a right-handed batter); batter fixed effects in that table absorb these level differences. $^{***}$ $p<0.01$, $^{**}$ $p<0.05$, $^{*}$ $p<0.1$ in two-sided $t$-tests with heteroskedasticity-robust standard errors.
\end{tablenotes}
\end{threeparttable}
\end{table}

\begin{table}[H]
\centering
\caption{\label{tab:robustness}Robustness of the main estimates}
\centering
\fontsize{8}{10}\selectfont
\begin{threeparttable}
\begin{tabular}[t]{>{\raggedright\arraybackslash}p{1.5in}cccccc}
\toprule
\multicolumn{1}{c}{ } & \multicolumn{3}{c}{Calls (Table \ref{tab:discretion})} & \multicolumn{1}{c}{Pitchers (Table \ref{tab:reallocation})} & \multicolumn{2}{c}{Batters (Table \ref{tab:batters})} \\
\cmidrule(l{3pt}r{3pt}){2-4} \cmidrule(l{3pt}r{3pt}){5-5} \cmidrule(l{3pt}r{3pt}){6-7}
Specification & Star & Hitter ahead & Two strikes & Left-handed $A$ & First stage & Reach two strikes\\
\midrule
Baseline & 1.15$^{***}$ & -5.47$^{***}$ & 6.90$^{***}$ & 0.058$^{***}$ & 0.75$^{***}$ & 0.57\\
 & (0.38) & (0.48) & (0.31) & (0.015) & (0.11) & (0.59)\\
Border band 0.12 ft & 1.27$^{***}$ & -5.21$^{***}$ & 7.40$^{***}$ &  &  & \\
 & (0.48) & (0.61) & (0.38) &  &  & \\
Border band 0.25 ft & 1.01$^{***}$ & -5.12$^{***}$ & 6.21$^{***}$ &  &  & \\
 & (0.31) & (0.40) & (0.26) &  &  & \\
Coarser location bins (6 cm) & 0.66 & -5.21$^{***}$ & 6.16$^{***}$ &  &  & \\
 & (0.43) & (0.56) & (0.35) &  &  & \\
Finer location bins (1.5 cm) & 1.17$^{***}$ & -5.50$^{***}$ & 7.13$^{***}$ &  &  & \\
 & (0.39) & (0.50) & (0.32) &  &  & \\
Logit (log-odds coefficient) & 0.02 & -0.26$^{***}$ & 0.89$^{***}$ &  &  & \\
 & (0.06) & (0.08) & (0.07) &  &  & \\
Star: top 10\% & 0.55 &  &  &  &  & \\
 & (0.50) &  &  &  &  & \\
Star: top 30\% & 1.19$^{***}$ &  &  &  &  & \\
 & (0.33) &  &  &  &  & \\
Star: fixed 2021 status & 0.81$^{*}$ &  &  &  &  & \\
 & (0.42) &  &  &  &  & \\
Excluding 2026 & 1.06$^{***}$ & -5.52$^{***}$ & 6.94$^{***}$ &  & 0.75$^{***}$ & 0.47\\
 & (0.40) & (0.50) & (0.32) &  & (0.11) & (0.56)\\
Cells with 100+ baseline pitches &  &  &  & 0.054$^{***}$ &  & \\
 &  &  &  & (0.015) &  & \\
Cells with 300+ baseline pitches &  &  &  & 0.055$^{***}$ &  & \\
 &  &  &  & (0.018) &  & \\
\bottomrule
\end{tabular}
\begin{tablenotes}[para]
\item \textit{Notes:} 
\item Each row re-estimates the main coefficients under one change to the baseline specification, and blank cells are not affected by that change. The first three columns are the star, hitter-ahead and two-strike interactions with the ABS in Table \ref{tab:discretion}, in percentage points (log-odds in the logit row); the fourth is the acceleration $A$ of left-handed pitchers in Table \ref{tab:reallocation}; and the last two are the first stage and the reduced-form effect of height on reaching two strikes in Table \ref{tab:batters}. Standard errors in parentheses below each estimate; $^{***}$ $p<0.01$, $^{**}$ $p<0.05$, $^{*}$ $p<0.1$ in two-sided tests with the same clustering as in the original tables.
\end{tablenotes}
\end{threeparttable}
\end{table}

\begin{table}[H]
\centering
\caption{\label{tab:fdr}Multiple testing: false-discovery-rate adjusted $q$-values}
\centering
\fontsize{9}{11}\selectfont
\begin{threeparttable}
\begin{tabular}[t]{llccc}
\toprule
Family & Test & Estimate & $p$ & $q$\\
\midrule
Calls (Table \ref{tab:discretion}) & Star $\times$ ABS & 1.149 & 0.003 & 0.004\\
Calls (Table \ref{tab:discretion}) & Star $\times$ ABS, batter fixed effects & 1.019 & 0.028 & 0.035\\
Calls (Table \ref{tab:discretion}) & Home $\times$ ABS & -0.123 & 0.634 & 0.634\\
Calls (Table \ref{tab:discretion}) & Hitter ahead $\times$ ABS & -5.468 & $<$0.001 & $<$0.001\\
Calls (Table \ref{tab:discretion}) & Two strikes $\times$ ABS & 6.901 & $<$0.001 & $<$0.001\\
Batters (Table \ref{tab:batters}) & High pitch thrown (reduced form) & 0.295 & 0.225 & 0.518\\
Batters (Table \ref{tab:batters}) & Swing at new strike (reduced form) & -1.248 & 0.652 & 0.652\\
Batters (Table \ref{tab:batters}) & Reach two strikes (reduced form) & 0.572 & 0.330 & 0.518\\
Batters (Table \ref{tab:batters}) & Reach three balls (reduced form) & -0.345 & 0.449 & 0.518\\
Batters (Table \ref{tab:batters}) & Two strikes, triple difference & 0.773 & 0.065 & 0.327\\
Batters (Table \ref{tab:batters}) & Hitter ahead, triple difference & 0.693 & 0.420 & 0.518\\
Batters (Table \ref{tab:batters}) & Star: walk & -0.324 & 0.466 & 0.518\\
Batters (Table \ref{tab:batters}) & Star: strikeout & 1.311 & 0.027 & 0.274\\
Batters (Table \ref{tab:batters}) & Star: on base & -0.886 & 0.179 & 0.518\\
Batters (Table \ref{tab:batters}) & Star: plate-appearance value & -5.682 & 0.401 & 0.518\\
Pitchers (Table \ref{tab:reallocation}) & Acceleration, All pitchers & 0.018 & 0.072 & 0.143\\
Pitchers (Table \ref{tab:reallocation}) & Acceleration, Left-handed & 0.058 & $<$0.001 & $<$0.001\\
Pitchers (Table \ref{tab:reallocation}) & Acceleration, Right-handed & 0.005 & 0.667 & 0.800\\
Pitchers (Table \ref{tab:reallocation}) & Acceleration, Side-arm & -0.007 & 0.826 & 0.826\\
Pitchers (Table \ref{tab:reallocation}) & Acceleration, Left-handed, balanced & 0.052 & 0.003 & 0.009\\
Pitchers (Table \ref{tab:reallocation}) & Acceleration, Right-handed, balanced & 0.009 & 0.432 & 0.648\\
\bottomrule
\end{tabular}
\begin{tablenotes}[para]
\item \textit{Notes:} 
\item Columns report the estimate, its two-sided $p$-value from the original table, and the Benjamini--Hochberg $q$-value computed within each family of tests (the rows of one table). A $q$-value below 0.05 means that the expected share of false discoveries among tests at least as significant is below 5\%.
\end{tablenotes}
\end{threeparttable}
\end{table}

\begin{table}[H]
\centering
\caption{\label{tab:regionevent}Called-strike probability by region and season, relative to 2023}
\centering
\fontsize{9}{11}\selectfont
\begin{threeparttable}
\begin{tabular}[t]{lccccc}
\toprule
\multicolumn{1}{c}{ } & \multicolumn{2}{c}{Human umpire} & \multicolumn{3}{c}{ABS} \\
\cmidrule(l{3pt}r{3pt}){2-3} \cmidrule(l{3pt}r{3pt}){4-6}
Region & 2021 & 2022 & 2024 & 2025 & 2026\\
\midrule
Zone, upper third & -16.1$^{***}$ (0.9) & -1.0 (0.7) & 13.0$^{***}$ (0.5) & 12.9$^{***}$ (0.5) & 12.9$^{***}$ (0.5)\\
High edge & -14.8$^{***}$ (0.8) & -4.3$^{***}$ (0.8) & 8.5$^{***}$ (0.8) & 0.5 (0.9) & -0.2 (0.9)\\
Low edge & 5.6$^{***}$ (1.1) & 4.6$^{***}$ (1.0) & -18.5$^{***}$ (0.8) & -9.9$^{***}$ (0.9) & -10.4$^{***}$ (1.0)\\
Inside edge & 2.3 (1.5) & 1.8 (1.4) & 8.2$^{***}$ (1.3) & 9.8$^{***}$ (1.5) & 9.5$^{***}$ (1.6)\\
Outside edge & 5.0$^{***}$ (1.2) & 9.8$^{***}$ (1.0) & -3.2$^{***}$ (1.0) & -2.7$^{**}$ (1.1) & -2.3$^{**}$ (1.1)\\
\bottomrule
\end{tabular}
\begin{tablenotes}[para]
\item \textit{Notes:} 
\item Columns report season coefficients relative to 2023, in percentage points, from linear probability models of a called strike for pitches in each region, with pitcher, batter and month fixed effects; regions are those of Table \ref{tab:geometry}. The 2021 column precedes the KBO's 2022 zone expansion. $^{***}$ $p<0.01$, $^{**}$ $p<0.05$, $^{*}$ $p<0.1$ in two-sided $t$-tests with standard errors clustered by game.
\end{tablenotes}
\end{threeparttable}
\end{table}

\begin{table}[H]
\centering
\caption{\label{tab:starbaseline}Star batters' outcomes under alternative baselines}
\centering
\begin{threeparttable}
\begin{tabular}[t]{lcc}
\toprule
Outcome & Baseline 2022--2023 & Baseline 2023 only\\
\midrule
Walk (pp) & -0.32 (0.44) & -0.59 (0.47)\\
Strikeout (pp) & 1.31$^{**}$ (0.59) & 1.89$^{**}$ (0.81)\\
On base (pp) & -0.89 (0.66) & -2.03$^{***}$ (0.70)\\
Plate-appearance value (wOBA points) & -5.68 (6.77) & -22.35$^{***}$ (8.02)\\
\bottomrule
\end{tabular}
\begin{tablenotes}[para]
\item \textit{Notes:} 
\item Columns report the star $\times$ ABS coefficient from Panel D of Table \ref{tab:batters}, with the pre-ABS baseline defined as 2022--2023 (as in the paper) or as 2023 alone; pp denotes percentage points, and wOBA (weighted on-base average) values each plate-appearance outcome by its average contribution to scoring. $^{***}$ $p<0.01$, $^{**}$ $p<0.05$, $^{*}$ $p<0.1$ in two-sided $t$-tests with standard errors clustered by batter.
\end{tablenotes}
\end{threeparttable}
\end{table}

\begin{table}[H]
\centering
\caption{\label{tab:lhp}Why did only left-handed pitchers accelerate?}
\centering
\fontsize{9}{11}\selectfont
\begin{threeparttable}
\begin{tabular}[t]{>{\raggedright\arraybackslash}p{0.95in}>{\raggedright\arraybackslash}p{2.2in}ccc}
\toprule
Explanation & Measure & Left-handed & Right-handed & Side-arm\\
\midrule
Shock size & Mean gain from ABS re-judging (wOBA points per 1,000 pitches) & 1.57 & 1.05 & 0.04\\
 & Share of pitches in gaining cells & 0.331 & 0.273 & 0.215\\
Timing & High-pitch share 2021 vs 2023 (pp) & -1.21 (1.05) & -1.93$^{***}$ (0.51) & -5.44$^{***}$ (1.24)\\
 & High-pitch share 2022 vs 2023 (pp) & -0.00 (0.76) & -1.52$^{***}$ (0.45) & -4.82$^{***}$ (1.11)\\
 & High-pitch share 2024 vs 2023 (pp) & 2.80$^{***}$ (0.78) & 3.93$^{***}$ (0.58) & 3.53$^{***}$ (1.09)\\
 & High-pitch share 2025 vs 2023 (pp) & 2.89$^{***}$ (0.79) & 2.51$^{***}$ (0.65) & 3.40$^{*}$ (1.93)\\
 & High-pitch share 2026 vs 2023 (pp) & 1.08 (0.90) & 1.61$^{***}$ (0.57) & 4.63$^{***}$ (1.51)\\
Matchup mix & Share of plate appearances against right-handed batters & 0.521 & 0.487 & 0.505\\
Pitcher type & Four-seam fastball velocity (km/h) & 142.6 & 144.9 & 139.9\\
 & Share of border pitches & 0.348 & 0.349 & 0.351\\
Composition & Acceleration $A$, balanced panel & 0.052$^{***}$ (0.018) & 0.009 (0.011) & \\
Human calls & High-zone strike calls 2022 vs 2023, relative to right-handed pitchers (pp) & -1.41 (0.90) & ref. & -3.74$^{***}$ (1.40)\\
\bottomrule
\end{tabular}
\begin{tablenotes}[para]
\item \textit{Notes:} 
\item Columns compare left-handed, right-handed and side-arm pitchers using 2022--2023 pitches unless stated, and rows group the measures by the candidate explanation they test; wOBA (weighted on-base average) values each plate-appearance outcome by its average contribution to scoring, and pp denotes percentage points. The timing rows are within-pitcher season coefficients of the high-pitch share (2023 omitted, standard errors clustered by pitcher), the acceleration row repeats Table \ref{tab:reallocation} for the balanced panel, the last row compares human high-zone calls across pitcher hands at the same location (standard errors clustered by game), and the remaining rows are descriptive. $^{***}$ $p<0.01$, $^{**}$ $p<0.05$, $^{*}$ $p<0.1$ in two-sided tests.
\end{tablenotes}
\end{threeparttable}
\end{table}

\begin{table}[H]
\centering
\caption{\label{tab:home}Home-team bias in calls}
\centering
\begin{threeparttable}
\begin{tabular}[t]{lccc}
\toprule
 & Estimate (pp) & SE & N\\
\midrule
Home batter, pre-ABS & 0.15 & (0.24) & 200,568\\
Home batter $\times$ ABS & -0.12 & (0.26) & 200,568\\
\bottomrule
\end{tabular}
\begin{tablenotes}[para]
\item \textit{Notes:} 
\item Columns report the coefficient in percentage points, its standard error and the number of border pitches, from a linear probability model of a called strike that compares home with away batters, with location-bin $\times$ season, pitcher, batter and home-plate-umpire fixed effects. $^{***}$ $p<0.01$, $^{**}$ $p<0.05$, $^{*}$ $p<0.1$ in two-sided $t$-tests with standard errors clustered by game.
\end{tablenotes}
\end{threeparttable}
\end{table}

\begin{figure}[H]\centering
  \includegraphics[width=\textwidth]{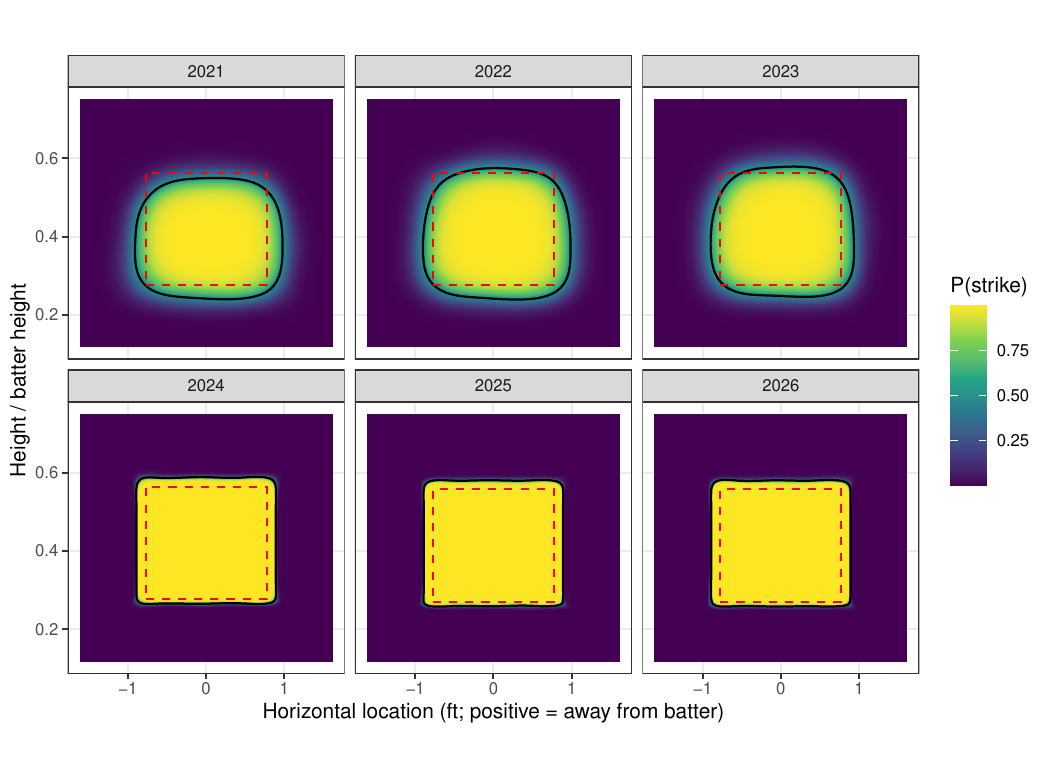}
  \caption{Called-strike probability by season}\label{fig:zonemaps}
  \begin{minipage}{0.95\textwidth}\footnotesize\textit{Notes:} Predicted probability of a called strike from a tensor-product
  logit smooth of horizontal location and height relative to the batter's registered height, estimated separately by season.
  Black lines are 50\% contours; red dashed boxes are the ABS zone (2024 rule through 2024, 2025 rule thereafter).
  2021 precedes the KBO's 2022 official zone expansion.\end{minipage}
\end{figure}

\begin{figure}[H]\centering
  \includegraphics[width=0.75\textwidth]{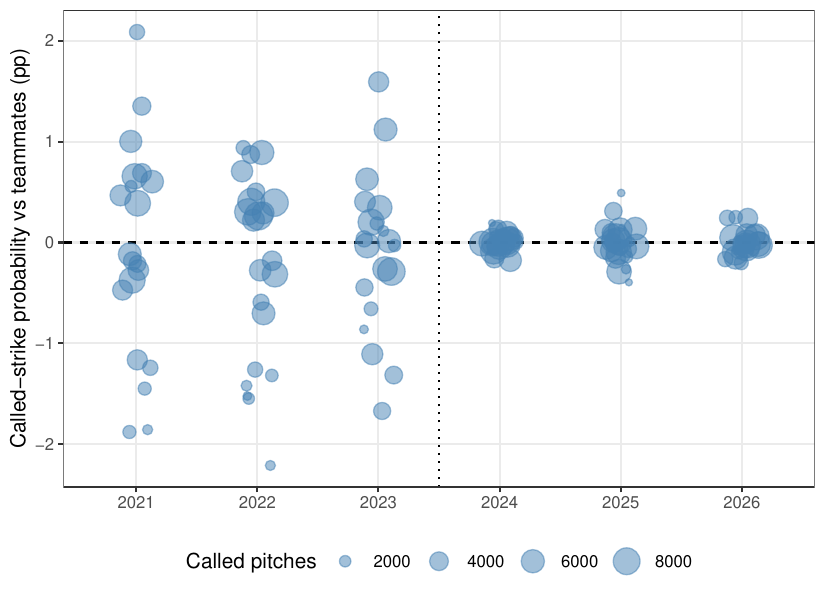}
  \caption{Catcher framing effects by season}\label{fig:framing}
  \begin{minipage}{0.95\textwidth}\footnotesize\textit{Notes:} Each point is a catcher-season estimate from Table~\ref{tab:framing},
  relative to teammates. Point size is the number of called pitches. The dotted line marks the introduction of the ABS.\end{minipage}
\end{figure}

\begin{figure}[H]\centering
  \includegraphics[width=0.85\textwidth]{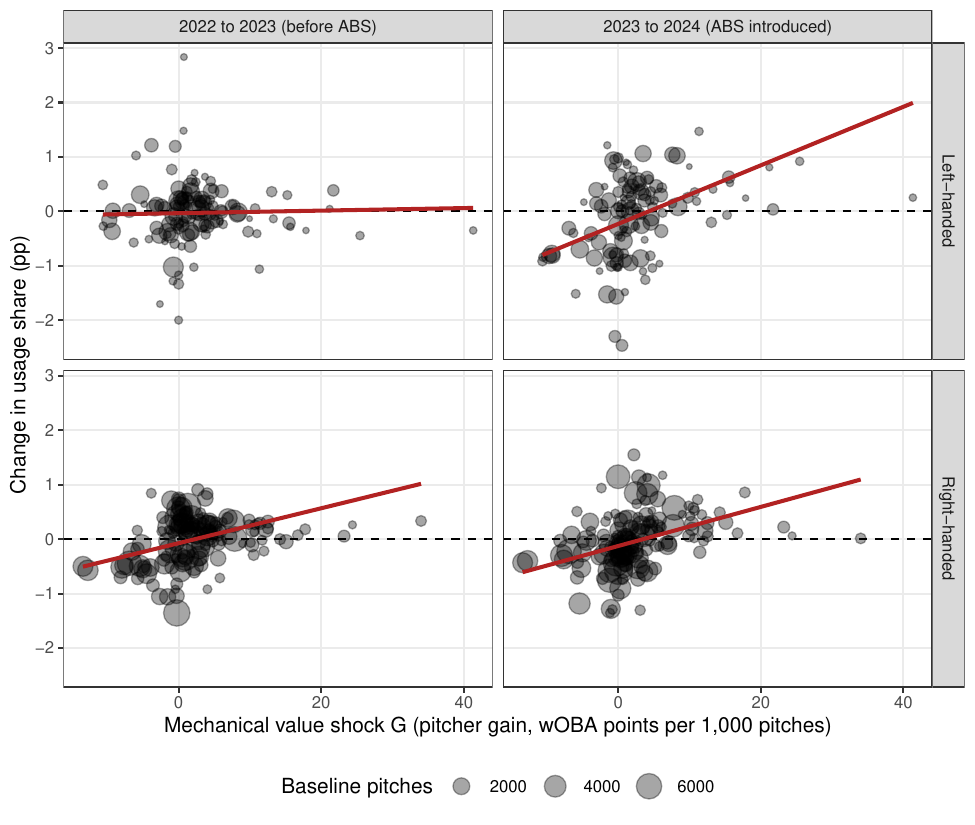}
  \caption{Mechanical value shock and the change in usage}\label{fig:shockresponse}
  \begin{minipage}{0.95\textwidth}\footnotesize\textit{Notes:} Each point is a cell (pitch type $\times$ vertical band $\times$ horizontal band).
  The horizontal axis is the cell's mechanical value shock $G$; the vertical axis is the change in its usage share.
  Lines are weighted least-squares fits, and pp denotes percentage points. See Table~\ref{tab:reallocation}.\end{minipage}
\end{figure}

\end{document}